\documentclass[12pt, preprint]{aastex}
\usepackage{natbib}
\shorttitle{Debris Disks in Kepler Exoplanet Systems }
\shortauthors{Lawler \& Gladman}

\newcommand{\Mearth}{\ensuremath{M_\oplus}}

\begin{document}

\title{Debris Disks in Kepler Exoplanet Systems}

\author{S.~M.~Lawler\altaffilmark{1}, B.~Gladman\altaffilmark{1}}
\altaffiltext{1}{University of British Columbia, Department of Physics and Astronomy, 6244 Agricultural Road, Vancouver, BC V6T 1Z1 Canada}

\begin{abstract}

The Kepler Mission recently identified 997 systems hosting
candidate extrasolar planets,
many of which are super-Earths.
Realizing these planetary systems are candidates to 
host extrasolar asteroid belts, we use mid-infrared data from the 
Wide-field Infrared Survey Explorer (WISE) to search for emission from dust
in these systems.
We find excesses around eight stars,
indicating the presence of warm to hot dust ($\sim$100-500~K), corresponding to
orbital distances of 0.1-10~AU for these solar-type stars.
The strongest detection, KOI~1099, demands $\sim$500~K dust interior to the
orbit of its exoplanet candidate.
One star, KOI~904, may host very hot dust ($\sim$1200~K, corresponding to
0.02~AU).
Although the fraction of these exoplanet-bearing stars with detectable warm excesses ($\sim$3\%) 
is similar to that found by Spitzer surveys of solar-type field stars,
the excesses detectable in the WISE data have much higher 
fractional luminosities (L$_{\rm dust}$/L$_{*}$) than
most known debris disks, implying that the fraction with debris disks of comparable luminosity may 
actually be significantly higher.
It is difficult to explain the presence of dust so close to the host stars,
generally corresponding to dust rings at radii $<$0.3~AU;
both the collisional and Poynting-Robertson drag timescales to remove dust from the system
are hundreds of years or less
at these distances.
Assuming a steady-state for these systems implies
large mass consumption rates with these short removal timescales,
meaning that
the dust production mechanism in these systems must almost certainly be episodic in nature.

\end{abstract}

\section{Introduction}

The evolution of solid circumstellar material is a competition between the accretional processes that
form planets (when speeds are low) and the violent collisional processes which in post-formation
systems inexorably grind down the material into particles small enough that radiation forces can
remove them from the system. 
The disappearance of infrared (IR) excesses around forming stars (on 3-10~Myr time scales) occurs because 
the accretional processes win and sequester the huge mass of dust into a (relatively) tiny number 
of objects with very low surface area/mass ratios. 
In contrast, the observed debris disks (known around hundreds of solar-type stars) are thought to be 
from collisional dust production as velocities rise during the final stages of planet assembly 
(for relatively young stars), or triggered around older stars as dust is liberated in higher-speed 
collisions \citep{Wyatt2008}. 
It is these more mature systems where our understanding of the physical picture is arguably incomplete. 
Although only $\sim$2\% of nearby main-sequence stars have massive detectable warm disks 
\citep[$>$200K;][]{Lawleretal2009}, several of these disks require dramatic hypotheses to produce the 
large dust mass estimated. 
That is, estimates of the collisional grinding of small-body belts indicate that the systems in 
question should not still possess dust at anywhere near the observed level \citep{Wyattetal2007}. 
In fact, \citet{Rheeetal2008} went so far as to liken the dust mass around BD~+20~307 as a `miracle' given
collisional-model predictions.

Debris disks are detected and studied via the excess IR flux they produce in the system,
caused by the dust re-radiating the
absorbed stellar light. 
This dust emission can then dominate the stellar photosphere in the IR, where the dust's blackbody peak is 
set by the equilibrium temperature at the orbital distance (for a narrow ring) or at an extended disk's 
inner edge, because the warmest dust usually dominates the systems dust emission. 
For `realistic' (although uncertain) dust grain size distributions, cross-sectional area 
and thus thermal emission is dominated by small grains, but the system mass is dominated by the 
unconstrained larger bodies. 
The true observable is thus the mass in grains in the decade larger than the peak wavelength.

Known disk systems consist of two types. 
The most common are massive and cold, analogous to the Solar System's Kuiper belt. 
A few hotter (T$>$200K) systems are known, which are massive cousins of our asteroid belt. 
An example is HD~69830, hosting 3 RV-discovered Neptune-mass planets inside 0.6 AU \citep{Lovisetal2006} 
and a warm dust ring at $\sim$1~AU, with an estimated 4x10$^{-7}$ M$_{\oplus}$ in small grains alone 
\citep{Lisseetal2007}. 
Production of this dust is problematic \citep{Beichman69830}, with hypotheses ranging from comet 
swarms to super-comets to massive planetary collisions; 
it is difficult to judge the plausibility of these mechanisms given the rarity of these outcomes. 

Dust grains in orbit around a star are quickly destroyed, and must be replenished by a 
fragmentation cascade, where asteroids
within a belt collide and break into progressively smaller pieces \citep{Wyatt2008}.
The lifetime of a debris disk will be the same as the lifetime of the largest bodies
in the source population that are available to be pulverized into dust grains.
Very small grains ($\lesssim$1~$\mu$m) are quickly blown away by stellar radiation pressure.
Larger grains suffer
Poynting-Robertson (PR) drag and spiral into the star in much less than the system's age;
for dust very close to the star ($\sim$0.1~AU), orbits collapse
in only hundreds of years. 

In this paper we search for excess infrared emission, indicating the presence of a dusty 
debris disk, in systems found by the Kepler mission to possess transiting exoplanets.
In Section~\ref{sec:WISE} we define our sample and discuss our methodology for 
identifying excess candidates, and 
in Section~\ref{sec:DDs} we provide additional details on the eight stars that have excesses.
Section~\ref{sec:dust} contains a discussion of the problems associated with having
this much dust so close to the host star, and finally Section~\ref{sec:conc} contains
a summary of recent work and this study's contribution relating debris disks to the 
presence of exoplanets.

\section{WISE Data} \label{sec:WISE}

WISE released preliminary survey data
in April 2011, covering about half the sky in photometric bands centered on 
3.4, 4.6, 12, and 22~$\mu$m \citep[W1, W2, W3, and W4, respectively;][]{Wrightetal2010}.  
Due to the pointing strategy of the telescope, the coverage on the sky is non-uniform.
Many individual images are stacked to produce the preliminary survey data, and some
areas of the sky were visited more often than others, providing deeper coverage
in those areas.

\subsection{Defining Our Sample}

We use WISE data to search for dust emission around the 928
solar-type stars (F, G, and K0-7 spectral types; T$_{\rm eff}$=4000-7200~K) 
that have stellar data (T$_{\rm eff}$, metallicity, and log($g$)) reported in the Kepler database
as of November 2011
and are candidates to host one or more planets 
contained in the first Kepler data release \citep{Boruckietal2011,Lissaueretal2011}.
Although most of these transiting planet candidates remain to be independently confirmed,
we will refer to them as planets for brevity.
The false-positive rate for single planet systems is $<$10\%
\citep{MortonJohnson2011}, and is much lower for multiplanet systems 
\citep{RagozzineHolman2010,LissauerNature2011}.
In any case, the existence or lack of planets is irrelevant
for our analysis of IR excess.

There are several M stars reported as having planets in the Kepler database, but we did not
consider these for analysis because of known systematic problems with modeling the atmospheres
of such cool stars \citep{Beichmanetal2006MIPS,Sinclairetal2010}.
We also did not consider the few hotter (A-type) stars with known Kepler planets in order to keep to
stars similar to our Sun.

The WISE preliminary data release covers about half the Kepler field, 
detecting 439 out of the 928 solar-type stars;
we consider a target to be `detected' in a given band if the WISE Source Catalog reports a 
signal-to-noise ratio (SNR) greater than 3.
(WISE does not report fluxes, but rather magnitudes and their corresponding magnitude uncertainty.)
All WISE detections are within 1~arcsecond of the coordinates given by the Kepler team.
The band wavelengths cover where hot dust emission would peak, which in these systems is also near
the equilibrium temperatures of the known Kepler planets, most of which are located 
0.01-0.4~AU from their host stars \citep{Lissaueretal2011}.

134 of these stars were flagged by the WISE team as having contamination from 
nearby, bright stars in the form of a halo or diffraction spike, or flagged as
extended, suggesting probable contamination by a background object.
These stars were not considered further.

The remaining 325 stars are all detected in both WISE's W1 (3.4~$\mu$m) and W2 (4.6~$\mu$m) bands, 
$\sim$2/3 of these are also are detected in W3 (12~$\mu$m), and only six are also detected in W4 (22~$\mu$m).

\subsection{Finding Excesses in the WISE Data}

To determine if the WISE data match the 
expected stellar photospheric spectrum or represent an excess of emission, we use 
Kurucz stellar atmosphere models \citep{CastelliKurucz2004} chosen from a grid of stellar 
temperatures, surface gravities, and metallicities to closely match those given for each 
star in the Kepler online database \citep{Brownetal2011}.
The Kurucz models are scaled using photometry from the Sloan Digital Sky Survey 
\citep[SDSS;][]{SDSS} and 
the Two Micron All-Sky Survey \citep[2MASS;][]{2MASS}.  
These Kepler targets are more distant than most known debris disk hosts (hundreds of parsecs), so extinction must be taken into
account.
The Kepler database provides $E(B-V)$ measurements for each star, which we scale for each photometric band 
\citep{RiekeLebofsky1985} and use to correct the photometry.
The Kurucz model is scaled to each star's flux using
the corrected photometry,
and then convolved with the efficiency function over each WISE passband.
The band zero points then provide expected magnitudes.

\begin{figure}[h]
\centering
\includegraphics[scale=0.6]{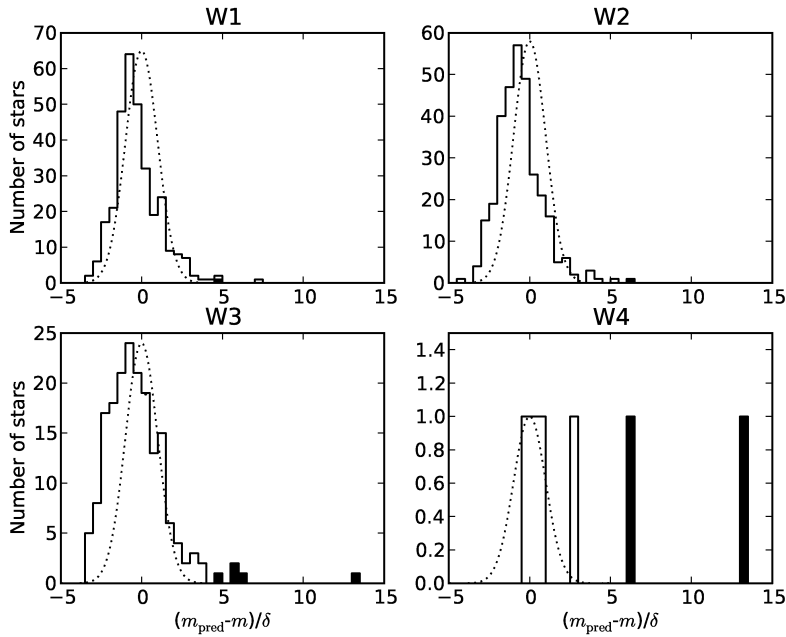}
\caption{
The excess significances for each band ($m_{\rm pred}$-$m$)/$\delta$. 
Overplotted are Gaussians with width $\delta$ for reference. 
Small ($\sim$1~$\delta$) systematics are still clearly present. 
Filled bars show what we accept as significant excesses. 
One system (KOI~379) shows excess in W1 and W2 above the 5 $\delta$ level 
(7 $\delta$ in W1 and 5 $\delta$ in W2) but is not considered to be significant because
of an obvious large flux mismatch between the SDSS and 2MASS photometry, perhaps indicating
stellar variability.
}
\label{fig:exhistos}
\end{figure}

\begin{figure}[h]
\centering
\includegraphics[scale=0.6]{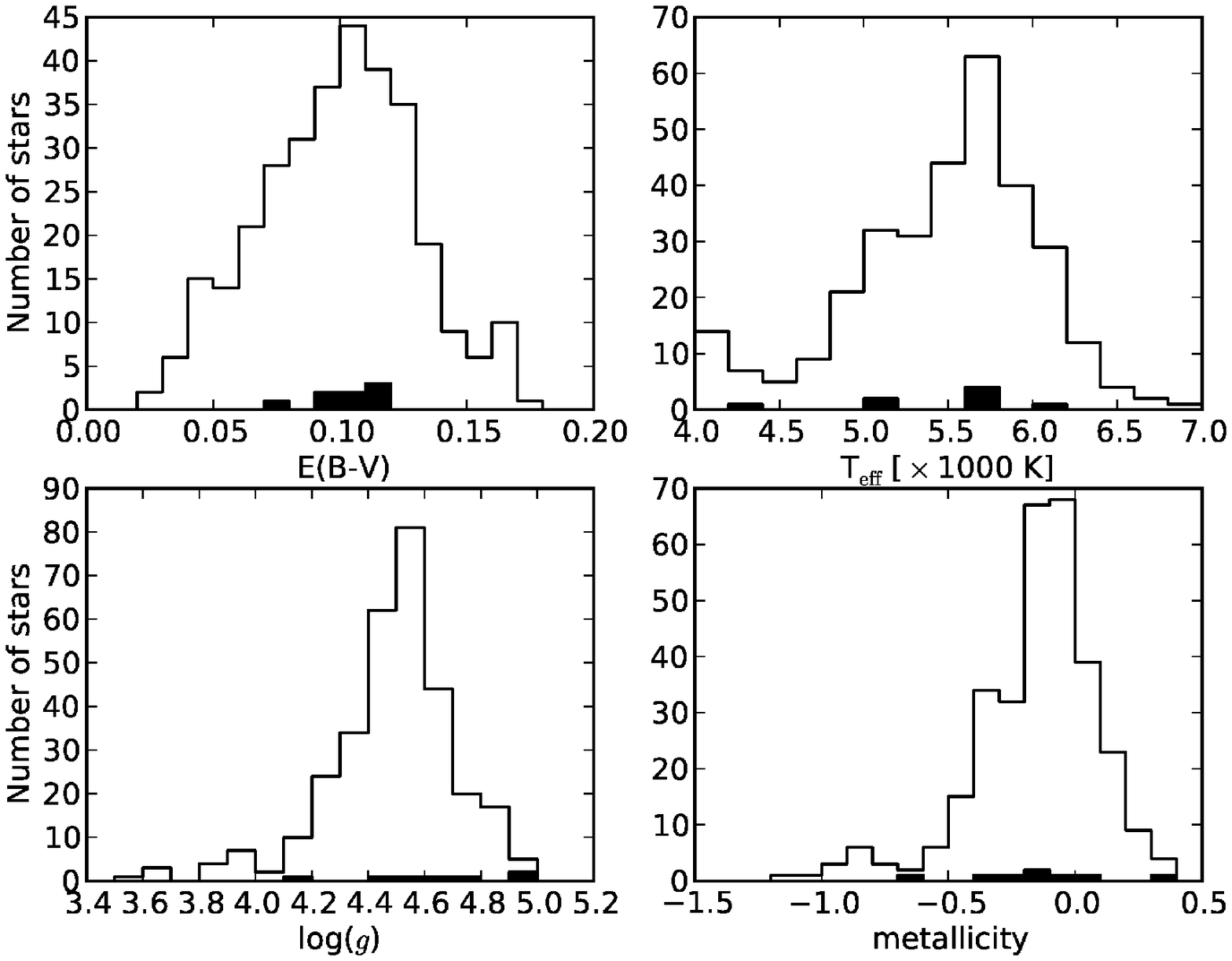}
\caption{
Histograms of each of the four stellar parameters reported by the
Kepler team.  
Filled bars show these values for the eight systems with significant
excess in at least one WISE band.
There appear to be no correlations between the presence of a significant
excess and any of these parameters.
}
\label{fig:datahistos}
\end{figure}

There are several sources of error in these measurements.  
To estimate magnitude uncertainties $\delta$, 
the given error bars for each datapoint ($\delta_W$) and
the WISE absolute photometric errors 
\citep[2.4, 2.8, 4.5, and 5.7\% in W1-4, respectively;][]{Wrightetal2010}
are added in quadrature.
To assess the significance of each possible excess,
the WISE magnitude ($m$) is subtracted from the predicted stellar magnitudes ($m_{\rm pred}$),
and this is divided by the uncertainty ($\delta$) in that band to give a dimensionless excess.
Our sign choice results in positive excesses indicating that the WISE flux is 
higher than the predicted flux.
The uncertainties in the stellar measurements reported by the Kepler database are
$\pm$200~K in temperature and 0.4~dex in log($g$) \citep{Brownetal2011}.  
We estimated the contribution of these uncertainties to the significance of the WISE excesses 
by using stellar models from (T$_{\rm eff}$, log($g$)) gridpoints that were varied by the error bars
in each of these parameters,
and we found that this resulted in $<$1~$\delta$ 
changes in the excess.
In order to account for this without having a formal number to add to our errors,
we adopt the fairly stringent
requirement that in order for an excess to be considered significant, it must be 
more than 5~$\delta$ above the photospheric magnitude.

Because of the large photometric aperture on the WISE measurements 
(8.25 arcseconds for W1-W3 and 16.5 arcseconds for W4), we visually inspected
both the WISE and SDSS images for each of the stars we find to have an excess
and found that there is not contamination from any nearby, bright stars.
In addition, we checked to make sure there were not systematic offsets in the 
expected position of the stars with excess, indicating contamination by a nearby 
background source.  

The distribution of excesses for each band are shown in Figure~\ref{fig:exhistos}.
They follow approximately Gaussian distributions, with very few datapoints falling above 5~$\delta$
and none falling below -5~$\delta$.
Figure~\ref{fig:exhistos} makes it clear that
the systematic error here actually underestimates the significance of the excesses; 
if anything we are being more conservative with our definition of what is a significant
excess and what is not.

Figure~\ref{fig:datahistos} shows stellar properties for our sample as reported by the Kepler team: 
extinction, stellar temperature, surface gravity, and metallicity.  
No trends are seen between any of these data and the presence of a significant excess in any band.

\section{Candidate Debris Disk Detections} \label{sec:DDs}

\begin{deluxetable}{ccccccccccccccc}																													
\tablecaption{FGK Stars with No Significant ($>$5~$\delta$) Excess\label{tab:noex}}																													
\tablehead{																													
\multicolumn{15}{c}{KOI}			}																										
\startdata																													
1	&	127	&	257	&	379	&	488	&	574	&	678	&	772	&	864	&	942	&	1094	&	1169	&	1288	&	1403	&	1489	\\
5	&	135	&	258	&	384	&	490	&	579	&	684	&	775	&	867	&	953	&	1095	&	1187	&	1302	&	1404	&	1501	\\
7	&	149	&	274	&	385	&	496	&	583	&	693	&	794	&	870	&	954	&	1101	&	1202	&	1308	&	1405	&	1503	\\
10	&	151	&	280	&	392	&	503	&	584	&	694	&	795	&	872	&	956	&	1102	&	1207	&	1309	&	1406	&	1505	\\
20	&	166	&	281	&	408	&	506	&	588	&	695	&	800	&	876	&	981	&	1108	&	1212	&	1310	&	1407	&	1515	\\
22	&	176	&	283	&	409	&	508	&	589	&	701	&	801	&	884	&	984	&	1109	&	1216	&	1311	&	1408	&	1519	\\
69	&	180	&	284	&	416	&	510	&	600	&	704	&	806	&	889	&	986	&	1110	&	1218	&	1312	&	1409	&	1521	\\
72	&	188	&	292	&	419	&	512	&	605	&	707	&	809	&	890	&	992	&	1115	&	1219	&	1314	&	1413	&	1525	\\
75	&	189	&	296	&	430	&	513	&	610	&	709	&	815	&	891	&	998	&	1116	&	1220	&	1337	&	1419	&	1526	\\
82	&	194	&	297	&	431	&	518	&	612	&	712	&	822	&	898	&	999	&	1117	&	1236	&	1338	&	1423	&	1529	\\
84	&	201	&	299	&	442	&	523	&	628	&	717	&	824	&	900	&	1001	&	1118	&	1240	&	1339	&	1427	&	1530	\\
98	&	202	&	301	&	443	&	524	&	632	&	718	&	830	&	901	&	1010	&	1128	&	1244	&	1360	&	1430	&	1533	\\
100	&	209	&	304	&	444	&	528	&	644	&	723	&	834	&	902	&	1022	&	1129	&	1261	&	1363	&	1432	&	1534	\\
103	&	221	&	313	&	454	&	530	&	647	&	732	&	837	&	911	&	1030	&	1141	&	1264	&	1366	&	1434	&	1541	\\
110	&	222	&	318	&	456	&	532	&	649	&	739	&	844	&	913	&	1031	&	1142	&	1266	&	1367	&	1436	&	1553	\\
111	&	223	&	339	&	457	&	533	&	650	&	740	&	847	&	918	&	1032	&	1148	&	1268	&	1370	&	1437	&	1587	\\
113	&	226	&	344	&	458	&	536	&	654	&	746	&	849	&	926	&	1052	&	1149	&	1273	&	1382	&	1445	&	1588	\\
115	&	227	&	345	&	472	&	537	&	661	&	755	&	853	&	928	&	1054	&	1150	&	1276	&	1387	&	1465	&	1605	\\
118	&	232	&	349	&	474	&	554	&	662	&	759	&	855	&	929	&	1061	&	1160	&	1278	&	1395	&	1477	&	1608	\\
122	&	241	&	351	&	475	&	557	&	663	&	765	&	858	&	934	&	1086	&	1168	&	1281	&	1396	&	1480	&	1609	\\
123	&	242	&	374	&	479	&	561	&	674	&		&		&		&		&		&		&		&		&		\\
\enddata																													
\end{deluxetable}

\begin{figure}[h!]
\centering
\includegraphics[scale=0.25]{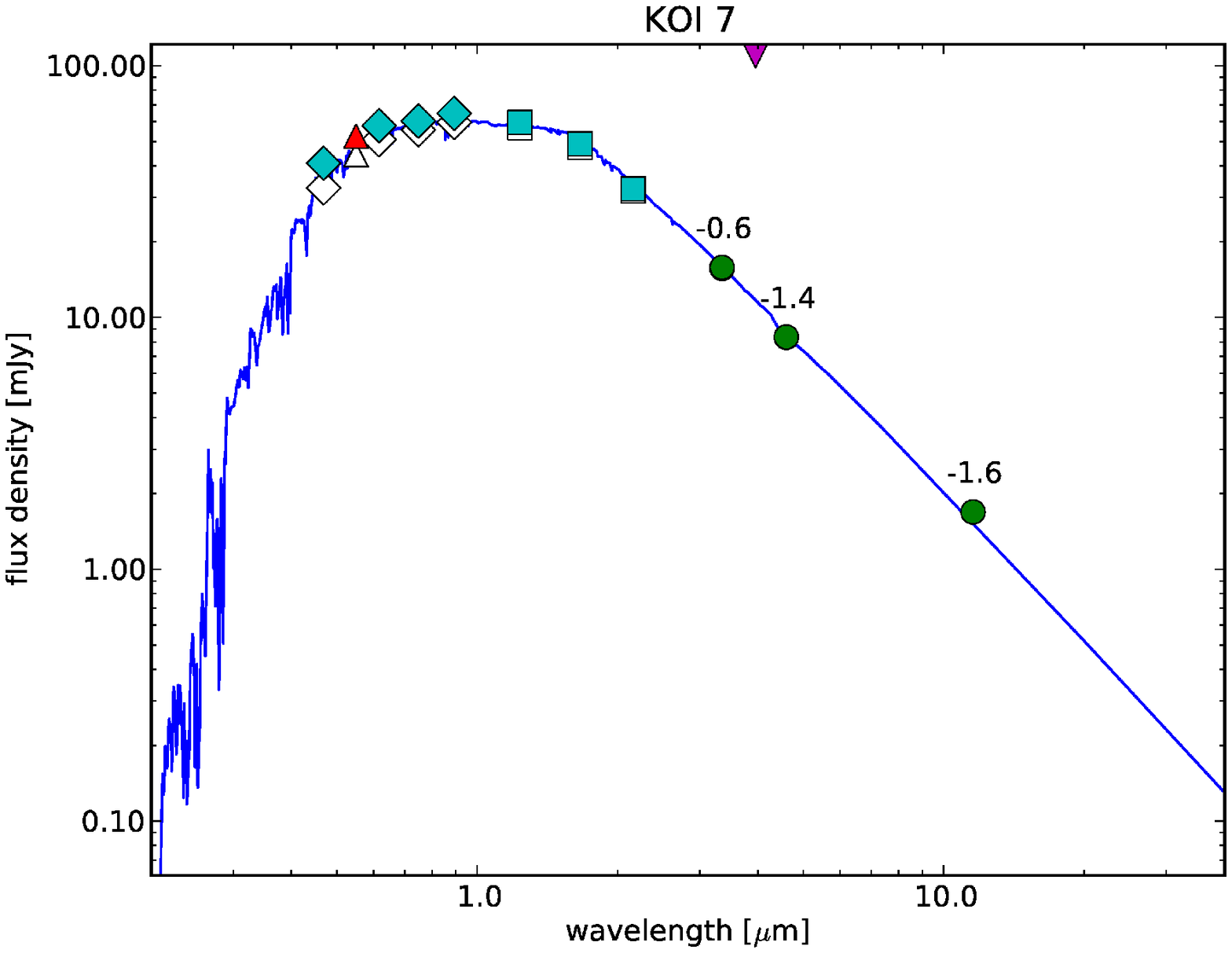} \includegraphics[scale=0.25]{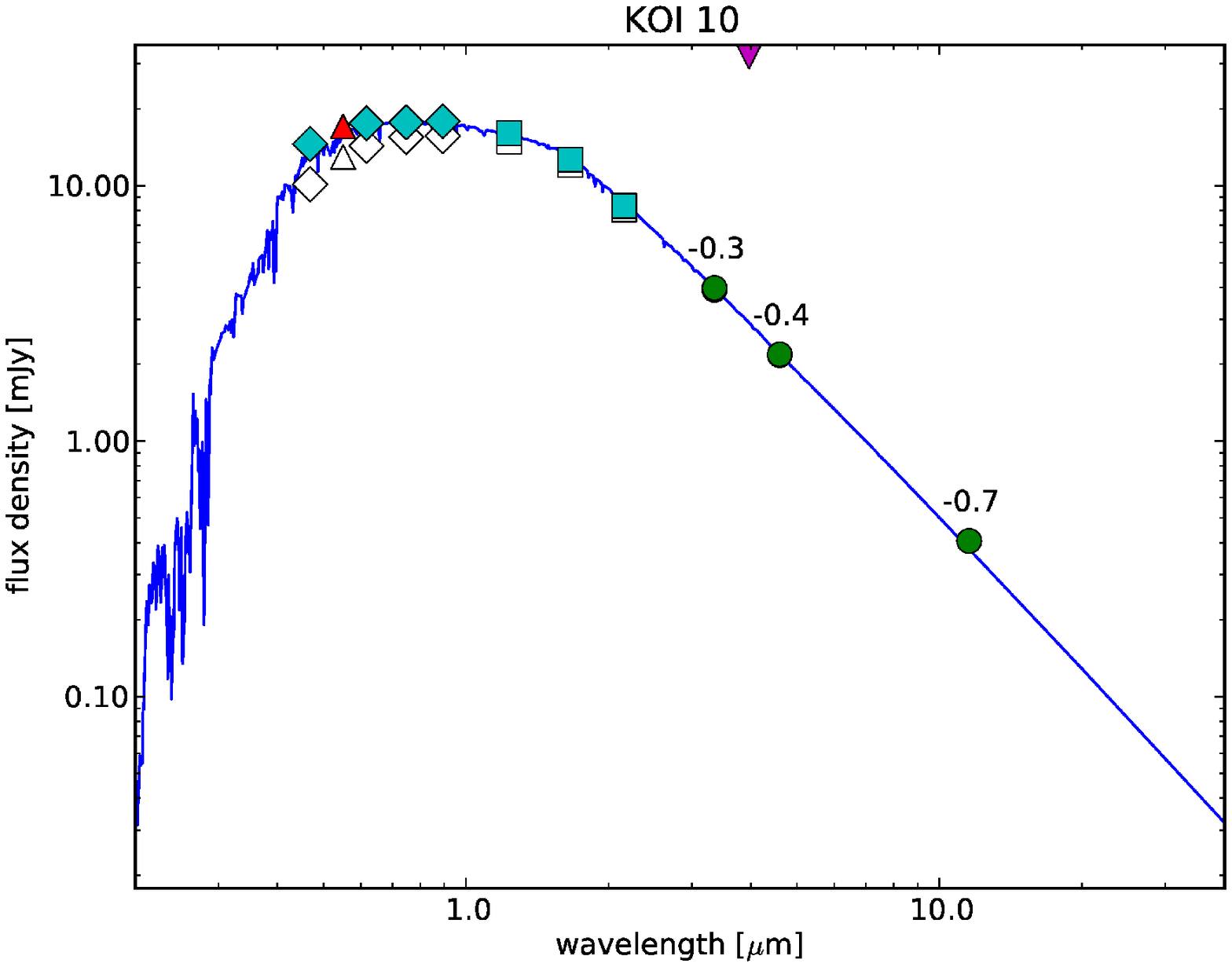} \includegraphics[scale=0.25]{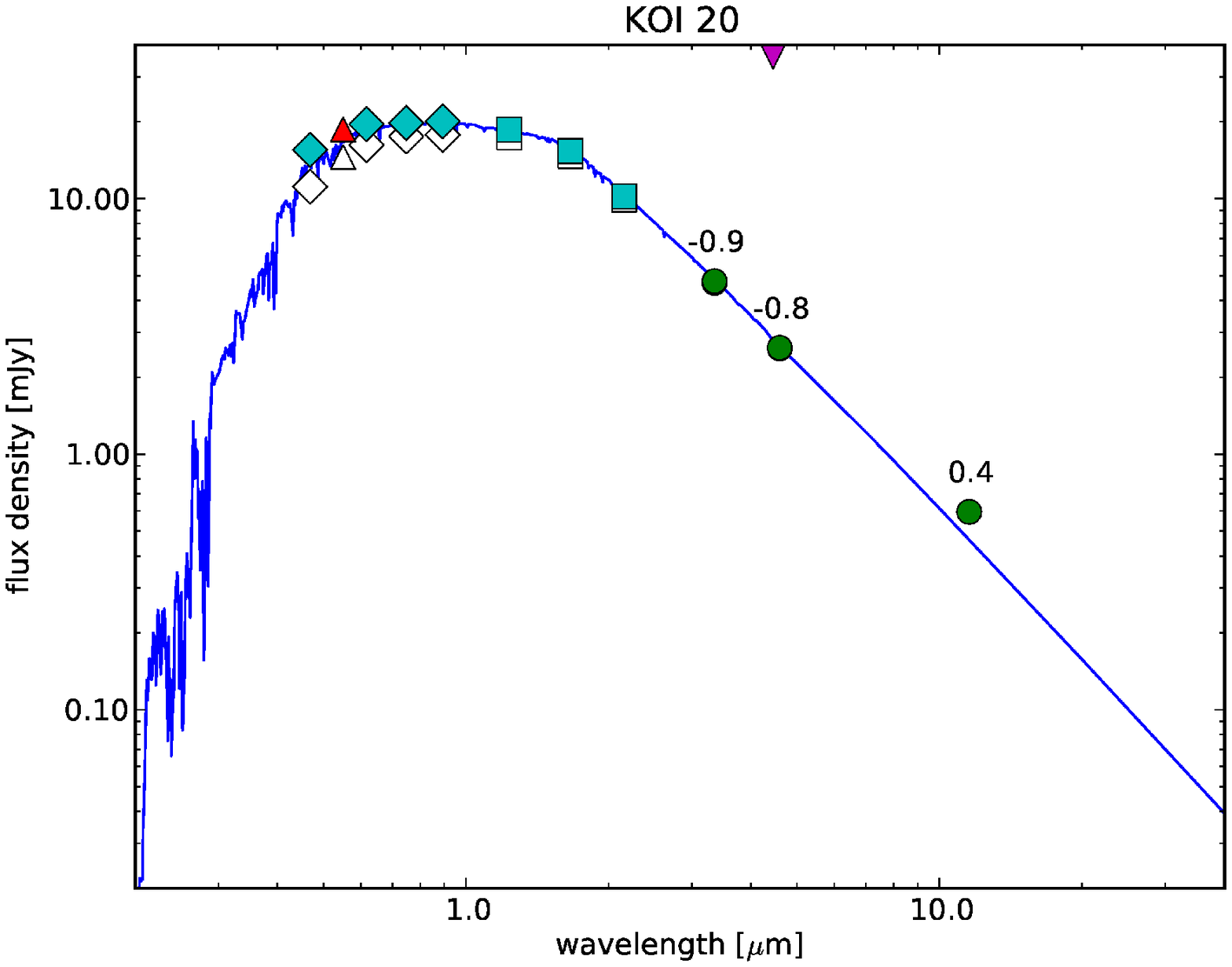}
\includegraphics[scale=0.25]{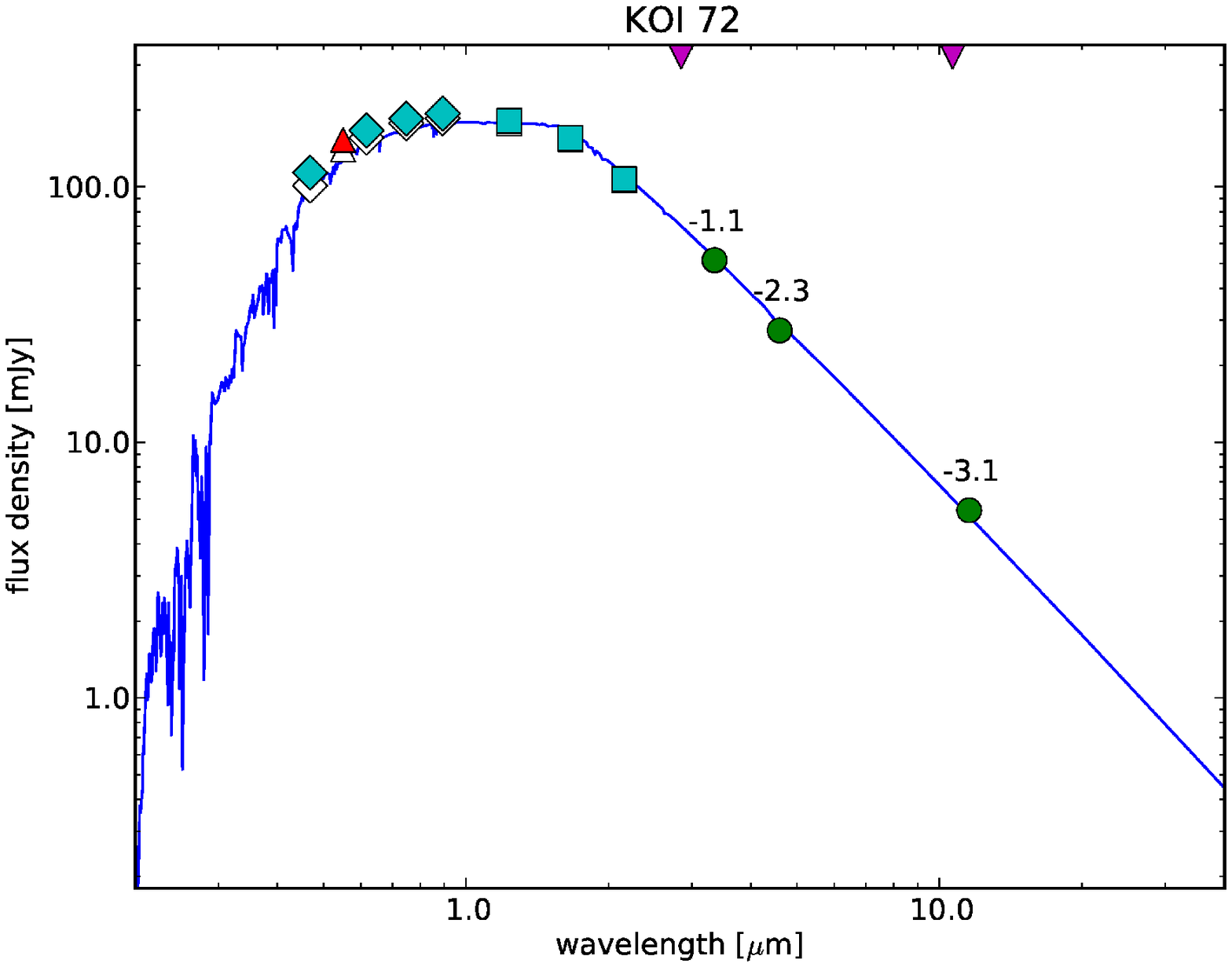} \includegraphics[scale=0.25]{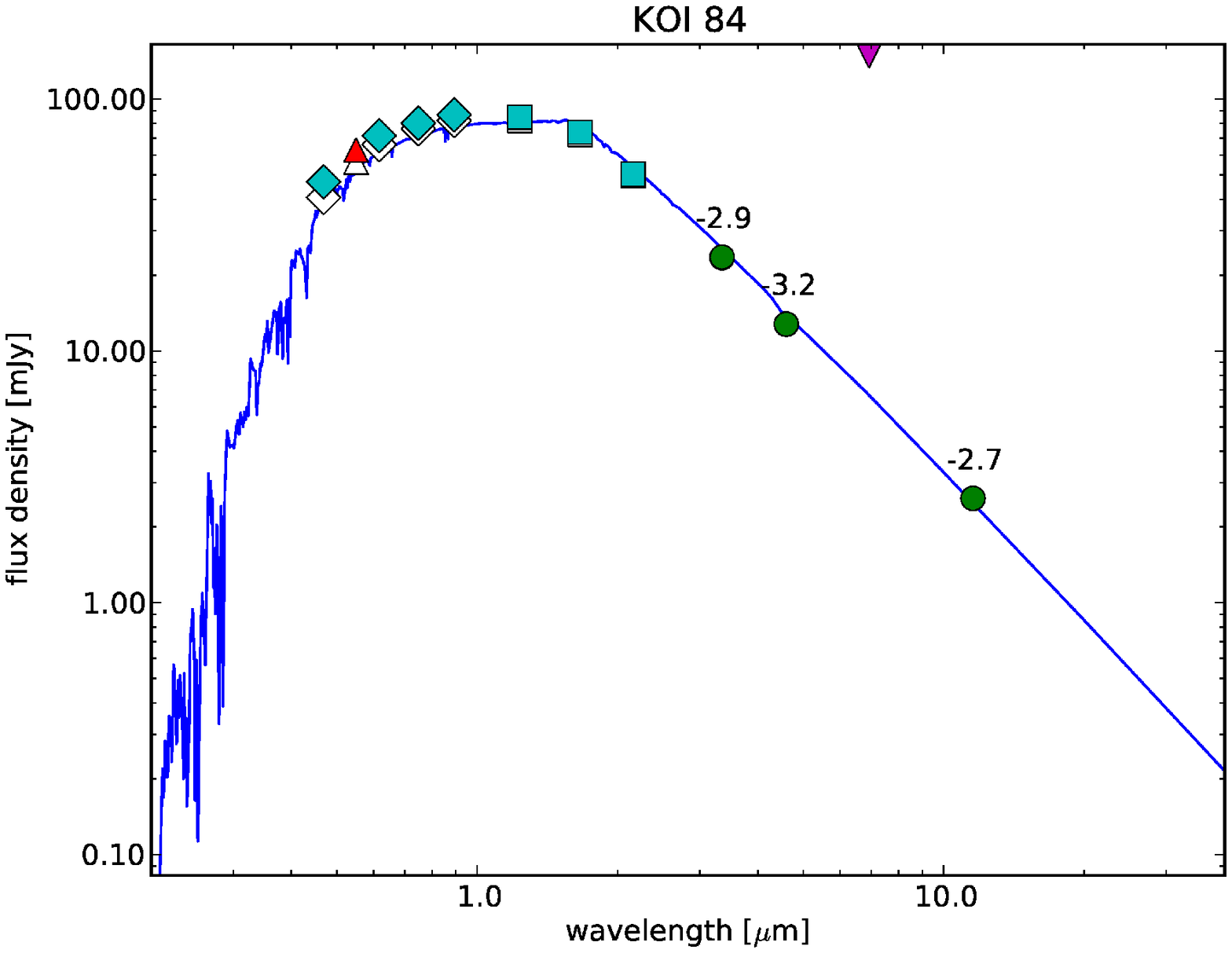} \includegraphics[scale=0.25]{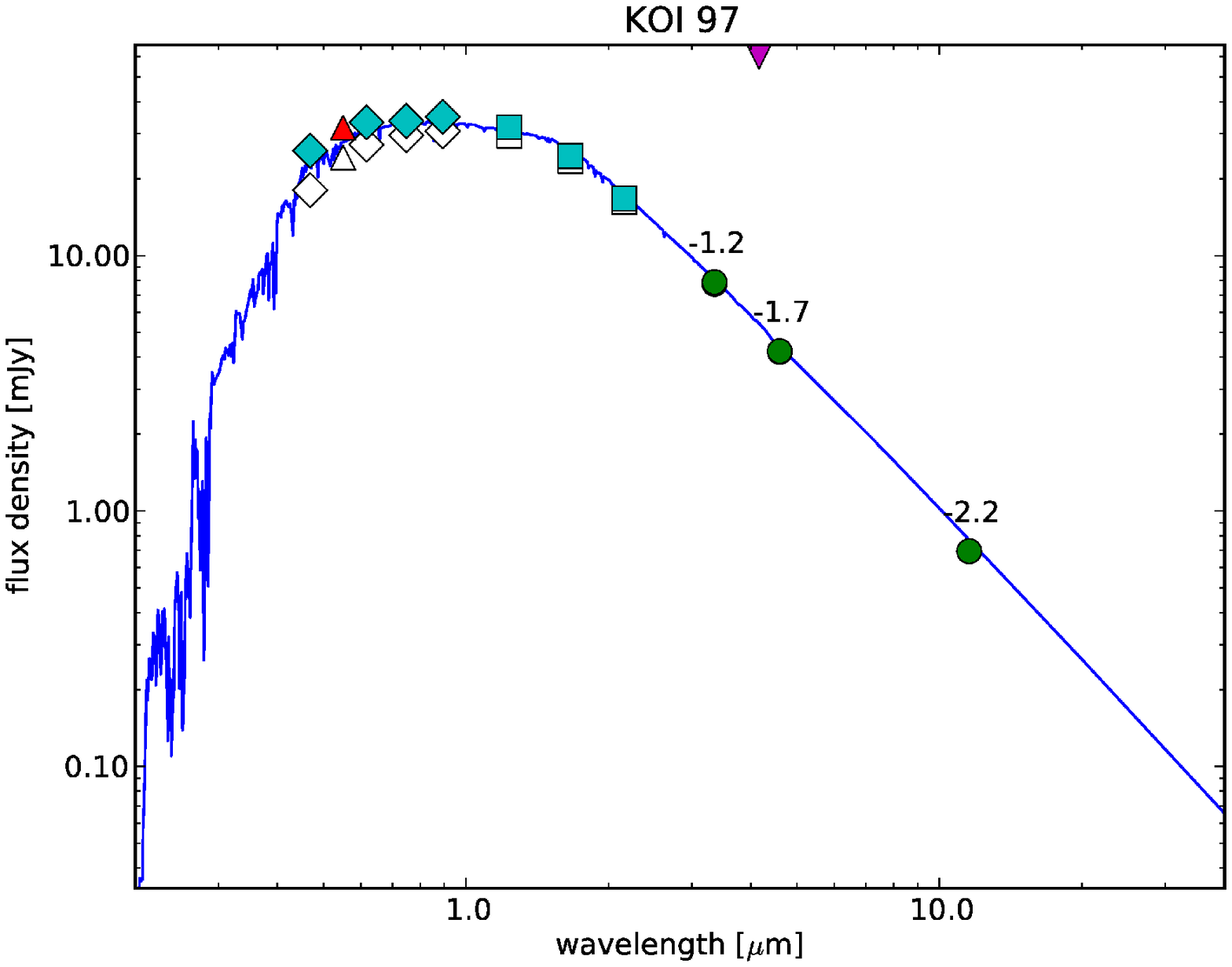}
\includegraphics[scale=0.25]{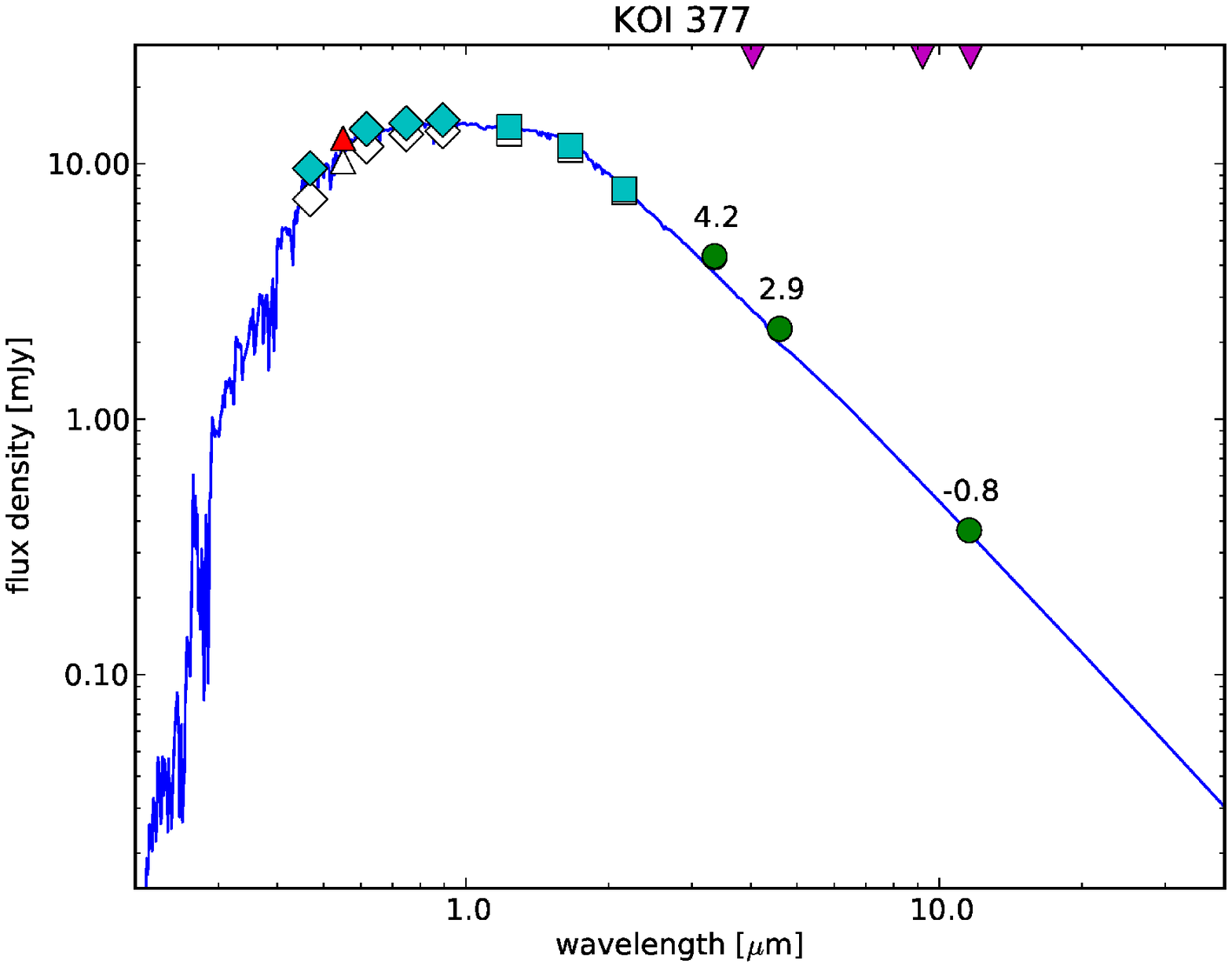} \includegraphics[scale=0.25]{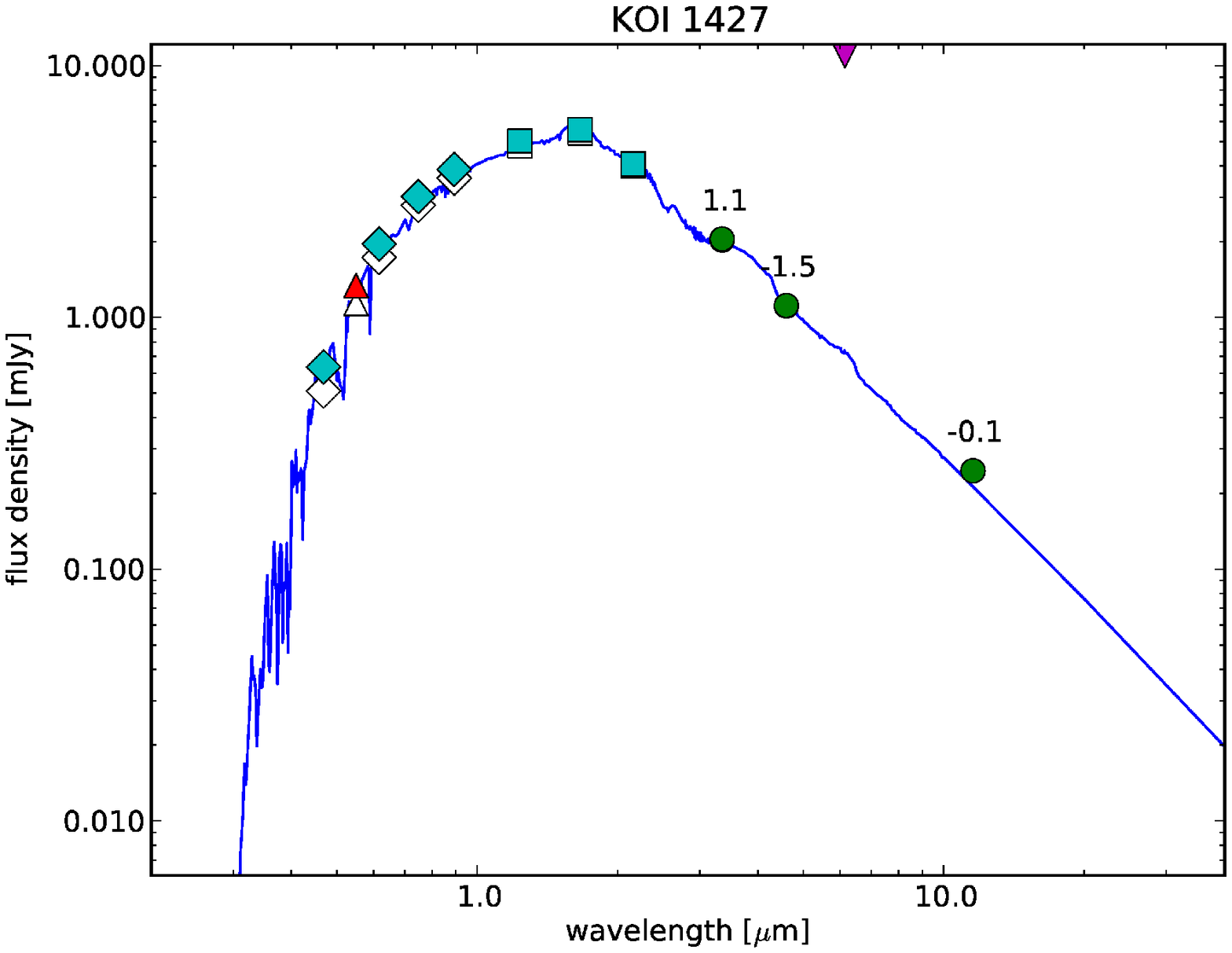} \includegraphics[scale=0.25]{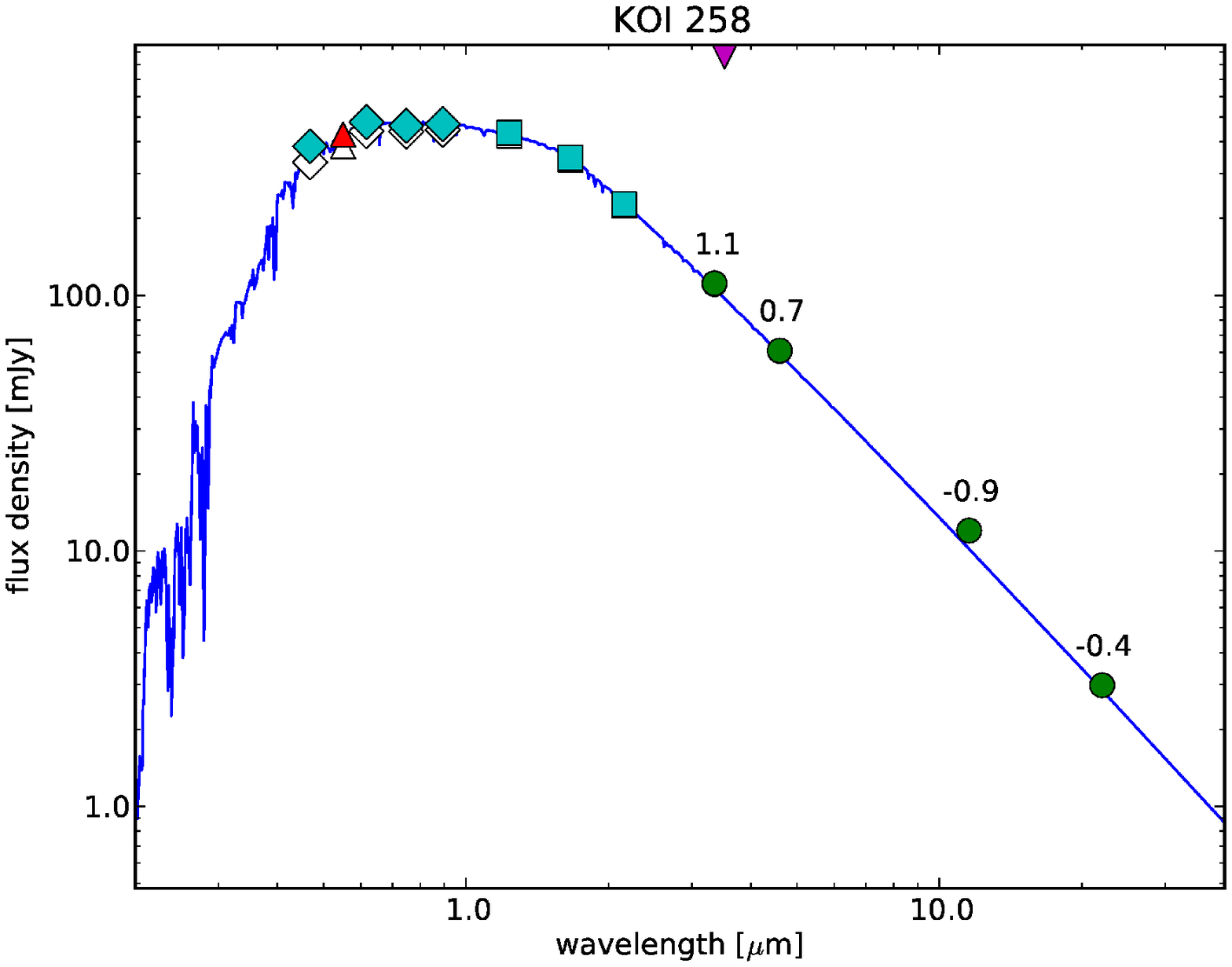}
\caption{
Nine stars with WISE data that is consistent with the stellar photosphere.
Included here are all of the confirmed systems that currently 
(as of November 2011) have WISE data 
(none of these systems have significant excesses):
KOI~7 = Kepler~4 \citep{Boruckietal2010Sci},
KOI~10 = Kepler~8 \citep{Boruckietal2010Sci},
KOI~20 = Kepler~12 \citep{Fortneyetal2011},
KOI~72 = Kepler~10 \citep{Batalhaetal2011},
KOI~84 = Kepler~19 \citep{Ballardetal2011},
KOI~97 = Kepler~7 \citep{Boruckietal2010Sci},
and KOI~377 = Kepler~9 \citep{Holmanetal2010}.
KOI~1427 and KOI~258 are included as additional examples, being one of the coolest and  
warmest stars in the sample, respectively.
Blue curve represents the stellar spectrum model.
Cyan diamonds and squares show the SDSS and 2MASS data respectively used to scale the stellar model,
and the red triangle shows the Kepler visual magnitude estimate converted to flux.
The filled symbols show the extinction-corrected data, and the empty symbols show the
original measured values.
Green filled circles show the WISE data, and 
the error bars are often smaller than the datapoints.
The excess (in multiples of $\delta$) of the observed-predicted magnitudes are given for the stellar model 
above each WISE datapoint.
At the top, triangles show the planetary thermal emission wavelength peak for reference.
}
\label{fig:noex}
\end{figure}

\begin{deluxetable}{l|cc|ccc|cccc|cccc}																																		
\rotate \tabletypesize{\scriptsize}																																		
\tablecaption{Debris Disk Candidates\label{tab:DDs}}																																		
\tablehead{																																		
	&	\multicolumn{5}{c}{Kepler Data}									&	\multicolumn{4}{c}{10~$\mu$m BB Dust Ring Model$^{a}$}											&	\multicolumn{4}{c}{1~$\mu$m Mod. BB Dust Ring Model$^{b}$}										\\
	&	\multicolumn{2}{c}{Stellar}			&	\multicolumn{3}{c}{Planetary}					&	\multicolumn{4}{c}{Dust Properties}											&	\multicolumn{4}{c}{Dust Properties}										\\
KOI	&	sp.	&	extinction	&	$a$	&	radius	&	mass$^d$	&	T		&	$a$	&	M				&	L$_{\rm d}$/L$_*$	&	T		&	$a$	&	M			&	L$_{\rm d}$/L$_*$	\\
	&	type$^c$	&	$E(B-V)$	&	[AU]	&	[$R_{\oplus}$]	&	[$M_{\oplus}$]	&	[K]		&	[AU]	&	[$\Mearth$]				&	[$\times$10$^{-3}$]	&	[K]		&	[AU]	&	[$\Mearth$]			&	[$\times$10$^{-3}$]	}
\startdata																																		
904$^e$	&	K5	&	0.074	&	0.029	&	2.1	&	5	&	1200$^f$	&	0.02	&	1$\times$10$^{-8}$		&	13	&	1100$^f$	&	0.02	&	2$\times$10$^{-9}$	&	15	\\
	&		&		&	0.159	&	3.0	&	10	&			&		&					&		&			&		&				&		\\ \hline
469	&	G0	&	0.116	&	0.095	&	5.5	&	35	&	400$^f$	&	0.36	&	8$\times$10$^{-7}$		&	3	&	300$^f$	&	0.64	&	2$\times$10$^{-7}$	&	3	\\
559	&	K0	&	0.099	&	0.052	&	1.4	&	2	&	471		&	0.23	&	4$\times$10$^{-7}$		&	4	&	382		&	0.34	&	1$\times$10$^{-7}$	&	4	\\
871	&	G5	&	0.092	&	0.105	&	10.9	&	144	&	329		&	0.27	&	6$\times$10$^{-7}$		&	4	&	283		&	0.37	&	1$\times$10$^{-7}$	&	4	\\
943	&	K0	&	0.102	&	0.045	&	2.2	&	5	&	357		&	0.28	&	1$\times$10$^{-6}$		&	6	&	303		&	0.39	&	2$\times$10$^{-7}$	&	6	\\
1099	&	G5	&	0.114	&	0.573	&	3.7	&	15	&	487		&	0.14	&	6$\times$10$^{-7}$		&	14	&	394		&	0.22	&	1$\times$10$^{-7}$	&	14	\\ \hline
1020	&	G3	&	0.11	&	0.294	&	21.9	&	614	&	100$^f$	&	9.89	&	5$\times$10$^{-4}$		&	3	&	90$^f$	&	12.21	&	6$\times$10$^{-5}$	&	2	\\
1564	&	G4	&	0.109	&	0.275	&	3.1	&	11	&	135$^f$	&	1.96	&	1$\times$10$^{-4}$		&	16	&	115$^f$	&	2.70	&	2$\times$10$^{-5}$	&	14	\\
\enddata																																		
\tablenotetext{a}{Assuming blackbody dust grains of radius 10~$\mu$m}																																		
\tablenotetext{b}{Assuming modified blackbody dust grains of radius 1~$\mu$m}																																		
\tablenotetext{c}{Spectral type is approximate and based on the stellar temperature provided by the Kepler team}																																		
\tablenotetext{d}{Planet mass is calculated according to $M/M_{\oplus}=(R/R_{\oplus})^{2.08}$ \citep{Boruckietal2011}}																																		
\tablenotetext{e}{KOI~904 has two known planets, both of which are listed here}																																		
\tablenotetext{f}{T$_{\rm dust}$ is assumed rather than fit.  See text for details on each system.}																																		
\end{deluxetable}

The vast majority of stars we examined have WISE fluxes consistent with the photosphere
(these are listed in Table~\ref{tab:noex} and a selection of these stars are shown in Figure~\ref{fig:noex}).
In our full sample, we find eight stars with an excess of 5~$\delta$ or higher in at least one band.
Unsurprisingly, these excesses generally appear in the longest two wavelength bands.
These stars are listed in Table~\ref{tab:DDs}, the WISE data for each of these systems are listed in Table~\ref{tab:WISE},
and we discuss each of them 
in Sections~\ref{sec:DDs:warm}-\ref{sec:DDs:hot} below.

Many simple debris disk models have been made using blackbody dust grains
\citep[i.e.][]{Beichmanetal2006,Hillenbrandetal2008,Lawleretal2009}.
We adopt 10~$\mu$m radius blackbody dust grains for an initial model, both for ease of comparison 
with previous literature and for simplicity given the lack of spectral data.
These models are not attempting to find the highly unconstrained `best' possible dust model
due to the myriad of complications inherent in dust emission 
(composition, radial distribution, emission features, and grain size), 
but rather to prove that the presence of dust provides a physically plausible model that 
approximately reproduces 
the WISE measurements.
A single temperature blackbody is scaled to fit the WISE photometry, giving L$_{\rm dust}$/L$_*$,
which is used to calculate the area of absorbing dust $\sigma_{\rm dust}$ using
$\sigma_{\rm dust}=4\pi a^2 {\rm L_{\rm dust}}/{\rm L_*}$
\citep{Wyatt2008}.
When a density and radius are assumed for the grains, this can be converted to the mass in dust.
We used 3.3~g~cm$^{-3}$ for silicate dust grains.
The temperature changes the wavelength of the peak of the curve.  
The orbital distance of the dust is then calculated using the same assumptions as
were used to calculate Kepler planet temperatures 
\citep[albedo of 0.3, uniform surface temperature, and 
no atmospheric effects;][]{Boruckietal2011}.

\begin{deluxetable}{l|cccc}																			
\tablecaption{WISE Data for Debris Disk Candidates\label{tab:WISE}}																			
\tablehead{																			
																			
KOI	&	W1 flux			&	W2 flux			&	W3 flux				&	W4 flux				\\
	&	[mJy]			&	[mJy]			&	[mJy]				&	[mJy]				}
\startdata																			
469	&	1.49	$\pm$	0.05	&	0.83	$\pm$	0.03	&	0.49	$\pm$	0.07		&		$<$	1.2		\\
559	&	2.31	$\pm$	0.07	&	1.28	$\pm$	0.04	&	0.60	$\pm$	0.08		&		$<$	2.5		\\
871	&	1.14	$\pm$	0.04	&	0.63	$\pm$	0.02	&	0.39	$\pm$	0.08		&		$<$	2.5		\\
904	&	1.94	$\pm$	0.06	&	1.03	$\pm$	0.04	&	0.22	$\pm$	0.07	$^a$	&		$<$	1.6		\\
943	&	0.97	$\pm$	0.03	&	0.51	$\pm$	0.02	&	0.36	$\pm$	0.07		&	1.5	$\pm$	0.5	$^a$	\\
1020	&	8.85	$\pm$	0.27	&	4.78	$\pm$	0.15	&	0.95	$\pm$	0.10		&	2.0	$\pm$	0.5		\\
1099	&	1.05	$\pm$	0.04	&	0.69	$\pm$	0.03	&	0.69	$\pm$	0.09		&	1.4	$\pm$	0.5	$^a$	\\
1564	&	1.11	$\pm$	0.04	&	0.59	$\pm$	0.02	&	0.29	$\pm$	0.07		&	2.0	$\pm$	0.6		\\
\enddata
\tablecomments{Fluxes and flux uncertainties are approximate, estimated from the given WISE magnitudes and magnitude errors.}																		
\tablenotetext{a}{SNR$<$3}																			
\end{deluxetable}

It is probable that smaller grains are present, 
and if they dominate in the size distribution of dust grains, 
the dust could be significantly farther out 
and produce the same fit.
We show this uncertainty in Table~\ref{tab:DDs}, where in addition to the best-fitting
10~$\mu$m dust grain model we show the best fit for a disk made of 1~$\mu$m modified blackbody
dust grains, where the emissivity of the grains drops proportional to $(\lambda/\lambda_0)^{-\beta}$,
with $\lambda_0$ equal to the grain size and $\beta$ equal to 1 \citep{Dentetal2000}.
This means that the spectral energy distrubution resulting from these grains is the same as a blackbody for 
wavelengths shorter than 1~$\mu$m, and drops off more steeply than a blackbody at larger wavelengths.
Smaller grains may be more realistic, as resolved disk studies have found larger
disks than predicted by fitting a blackbody spectrum to the mid-IR excess \citep[i.e.][]{Kristetal2010}.
However, in the interest of picking a uniform model that can be compared with
previous Spitzer studies, we use the 10~$\mu$m blackbody dust model for most of
our calculations.

For our 10~$\mu$m dust grain model,
the resulting blackbody curve is added to the stellar atmosphere model, and this is used
to fit the WISE data.
For most systems, the temperature and surface area of the emitting dust are both free parameters in the 
fit. 
Because in some systems there is only one datapoint that rises off the photosphere,
the temperature cannot be determined uniquely even if an upper limit can be set.
These systems are noted in Table~\ref{tab:DDs} and discussed below.
For all eight systems, the fit for temperature is fairly unconstrained;
the longest wavelength datapoint is almost always still rising, allowing a large
number of cooler blackbody curves from more massive dust disks to pass through the datapoints.
Thus the temperatures we calculate should be thought of as
upper limits, as a larger amount of somewhat cooler dust could also produce the observed emission.

\begin{figure}
\centering
\includegraphics[scale=0.35]{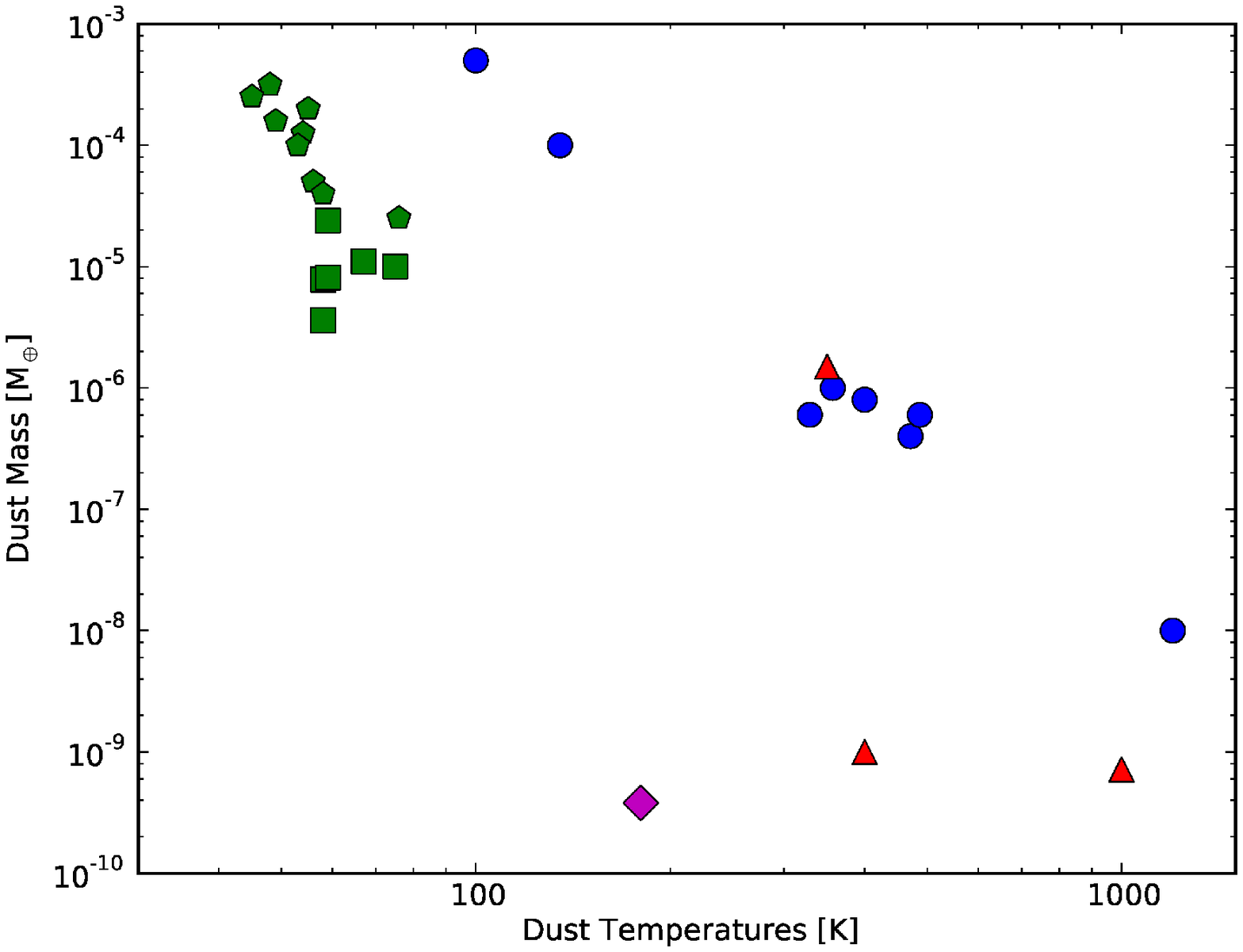} \includegraphics[scale=0.35]{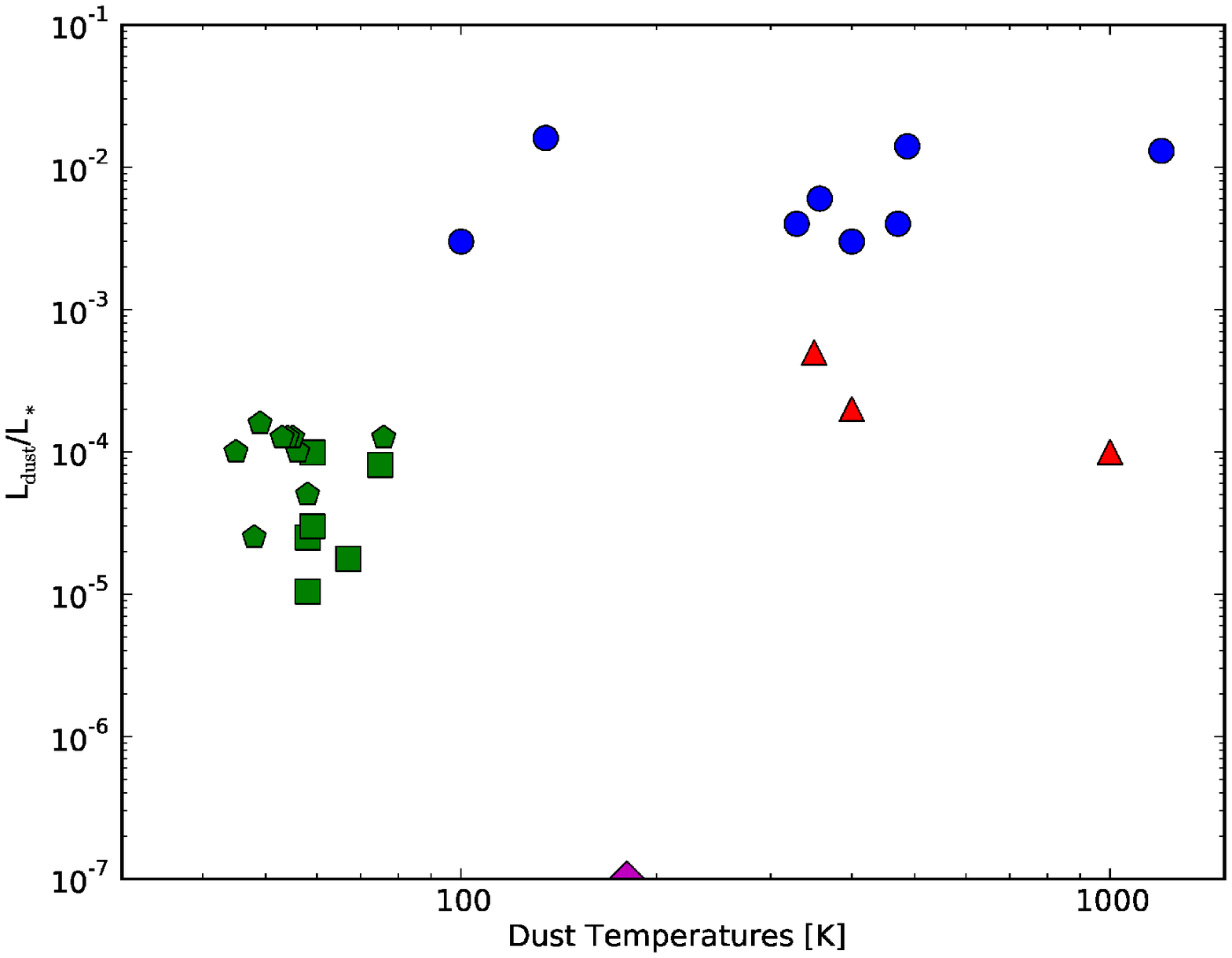}
\caption{
The derived masses, fractional luminosities, and temperatures of dust rings in our sample 
(using 10~$\mu$m radius grains) compared to 
other known systems.
Red triangles show systems examined by \citet{Wyattetal2007}
\citep[see also][]{Beichman69830,Beichmanetal2006,Lisseetal2011},
green squares are from a Spitzer spectroscopic survey \citep{Lawleretal2009}, 
and green pentagons are from a Spitzer photometric survey at 70~$\mu$m \citep{Hillenbrandetal2008}.
For comparison, the purple diamond shows 
zodiacal dust in our solar system \citep{Hahnetal2002}.
All systems shown are mature, solar-type stars.
Blue circles show the dust models presented in this paper.
Because $M_{\rm grain}$ is proportional to grain size $r_{\rm grain}$,
using smaller grains would result in lower total dust masses (see Table~\ref{tab:DDs}).
While the masses shown here are similar to known systems, because of the warmer temperatures we find for 
these Kepler systems, the resulting fractional luminosities L$_{\rm dust}$/L$_*$ are much higher.
}
\label{fig:dustlocations}
\end{figure}

Using this simple dust model,
the dust masses needed to produce these excesses range from $\sim$10$^{-8}$-10$^{-4}$ 
Earth masses, orders of magnitudes larger than the dust mass of 
our solar system \citep[$\sim4\times10^{-10}$~$\Mearth$;][]{Hahnetal2002},
and comparable in mass to known debris disk systems 
\citep[typically 10$^{-8}$$-$10$^{-2}$~$\Mearth$;][]
{Beichman69830,Brydenetal2006,Wyattetal2007,Rheeetal2008,Hillenbrandetal2008},
but with higher fractional luminosities
as shown in Figure~\ref{fig:dustlocations}.

We find that our eight excess systems can fit into three categories based on dust temperature
(and thus dust location).
For much of the literature, $>$200~K dust is considered to be hot dust.
Due to the large variation in debris disk temperatures and terminology, 
for this paper we define our debris disk categories as the following:
hot ($\sim$1000~K), warm ($\sim$300-500~K), and cool ($\sim$100~K).
We now discuss each excess system in turn.

\subsection{Warm Dust Systems ($\sim$300-500~K)} \label{sec:DDs:warm}

\begin{figure}[h]
\centering
\includegraphics[scale=0.6]{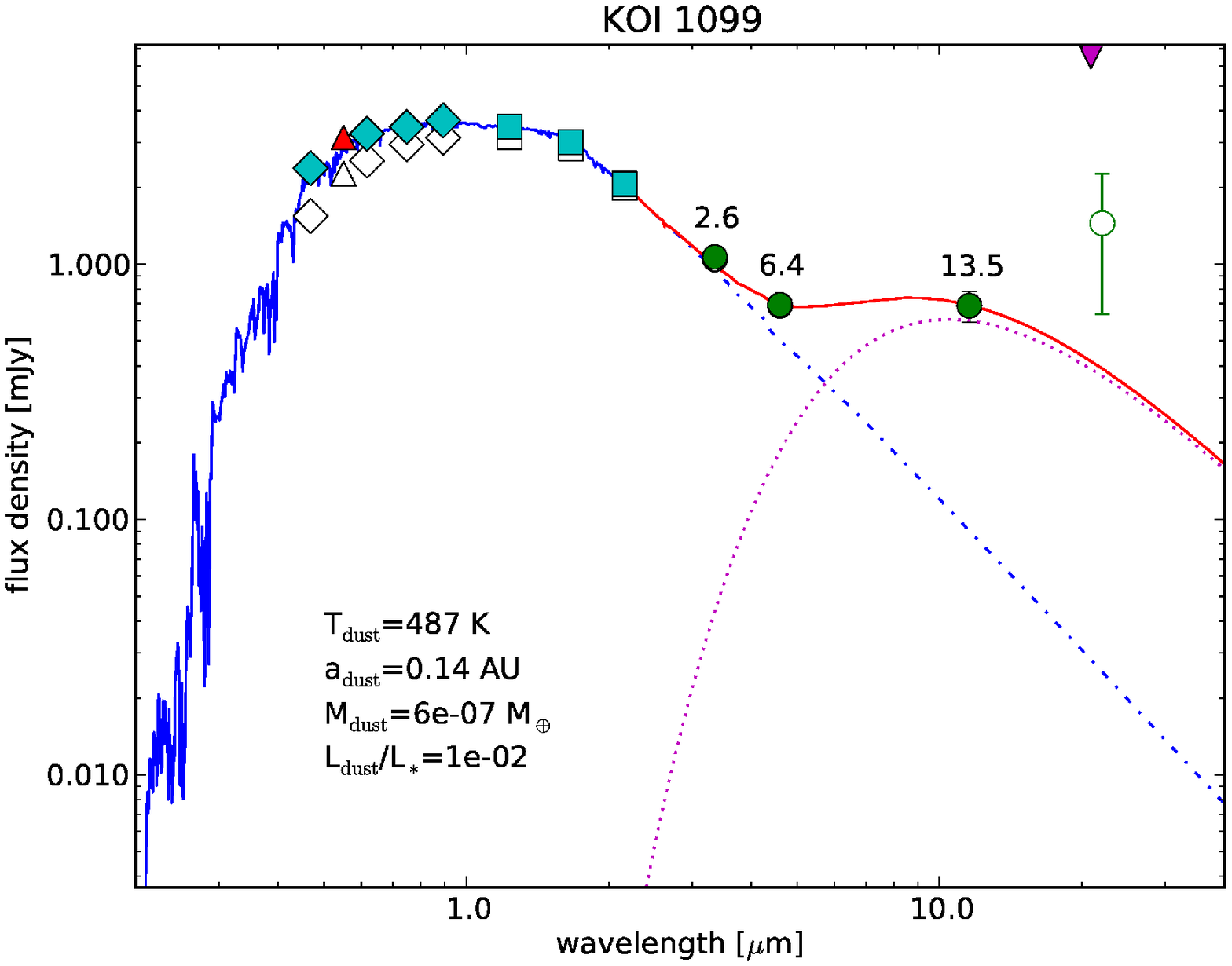}
\caption{
KOI~1099, our best candidate for a debris disk due to the
strong excess in W2 and W3, and a weak excess in W1.
See caption for Figure~\ref{fig:noex}.
The excess values (in multiples of the total magnitude uncertainty $\delta$) 
for the stellar atmosphere model predictions (not including the dust
model) are shown above each WISE datapoint.
Additionally, the purple dotted curve shows the blackbody spectrum of a thin ring of
10~$\mu$m-radius dust grains, and the solid red curve shows the star+dust spectrum.
The best-fitting dust mass, temperature, and orbital distance are also shown.
The unfilled WISE datapoint shows SNR$<$3 data that was not used in the fit,
but gives a preview of where the upcoming WISE full data release may provide
significant data.
}
\label{fig:sed1099}
\end{figure}

\begin{figure}[h]
\centering
\includegraphics[scale=0.4]{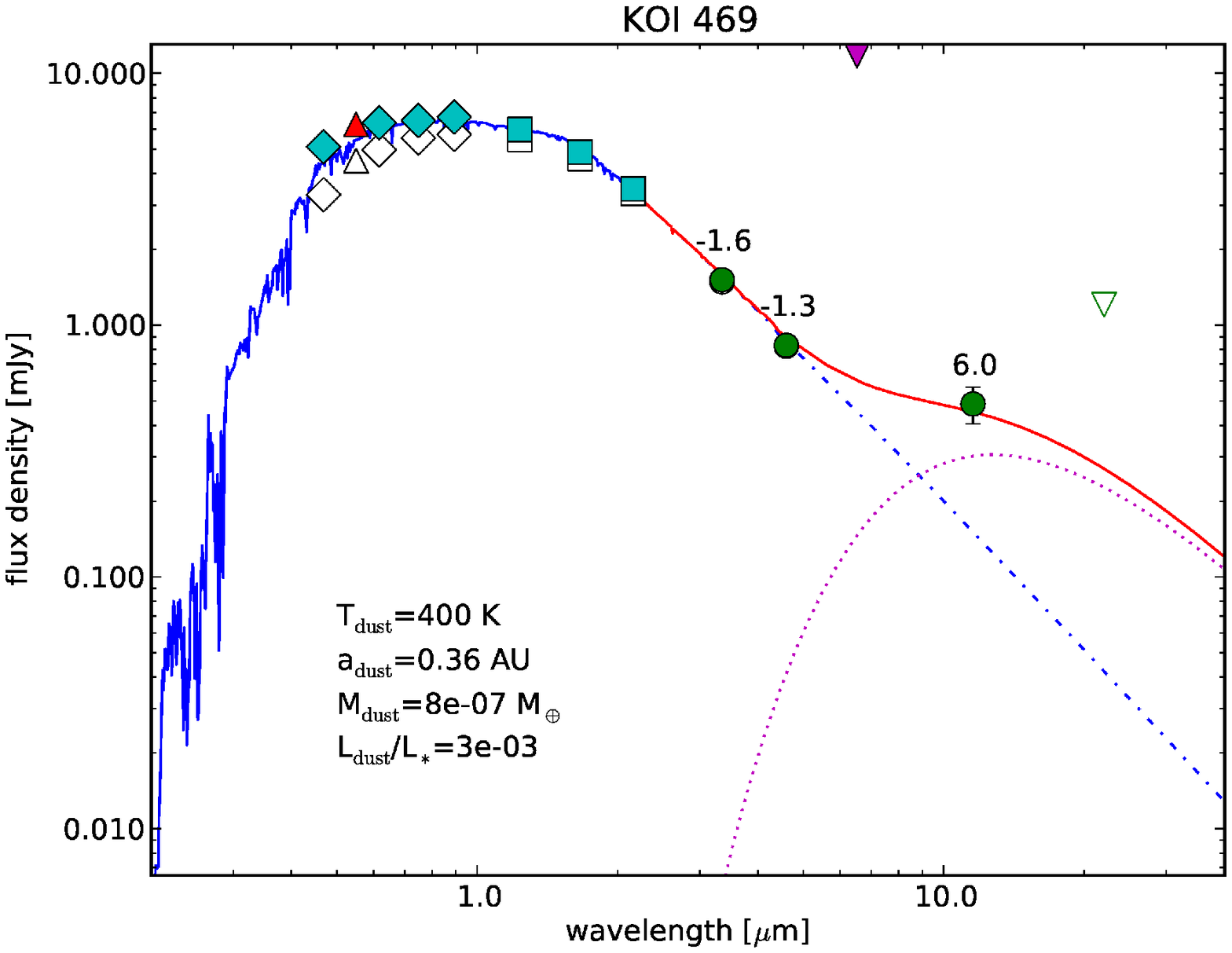} \includegraphics[scale=0.4]{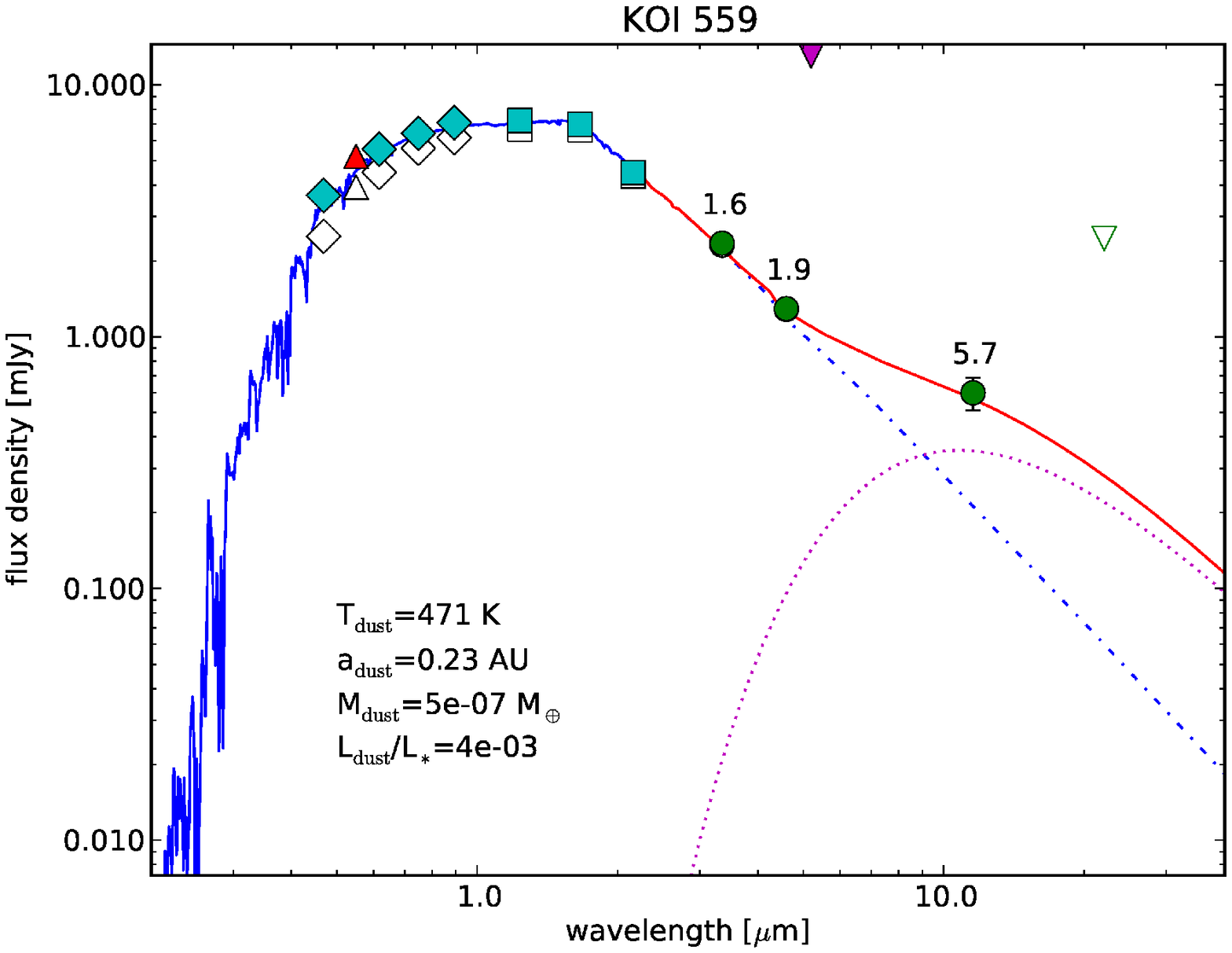}
\includegraphics[scale=0.4]{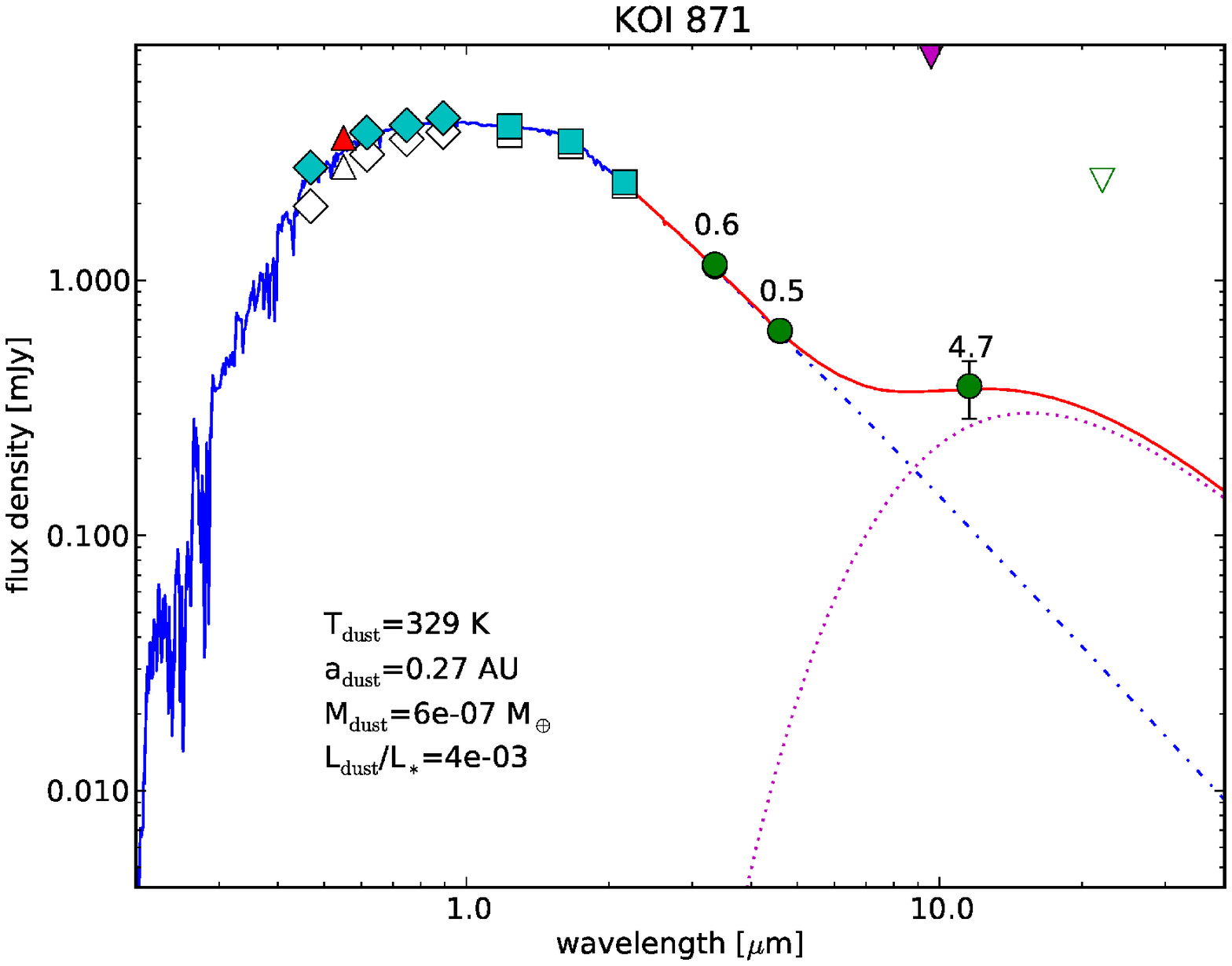} \includegraphics[scale=0.4]{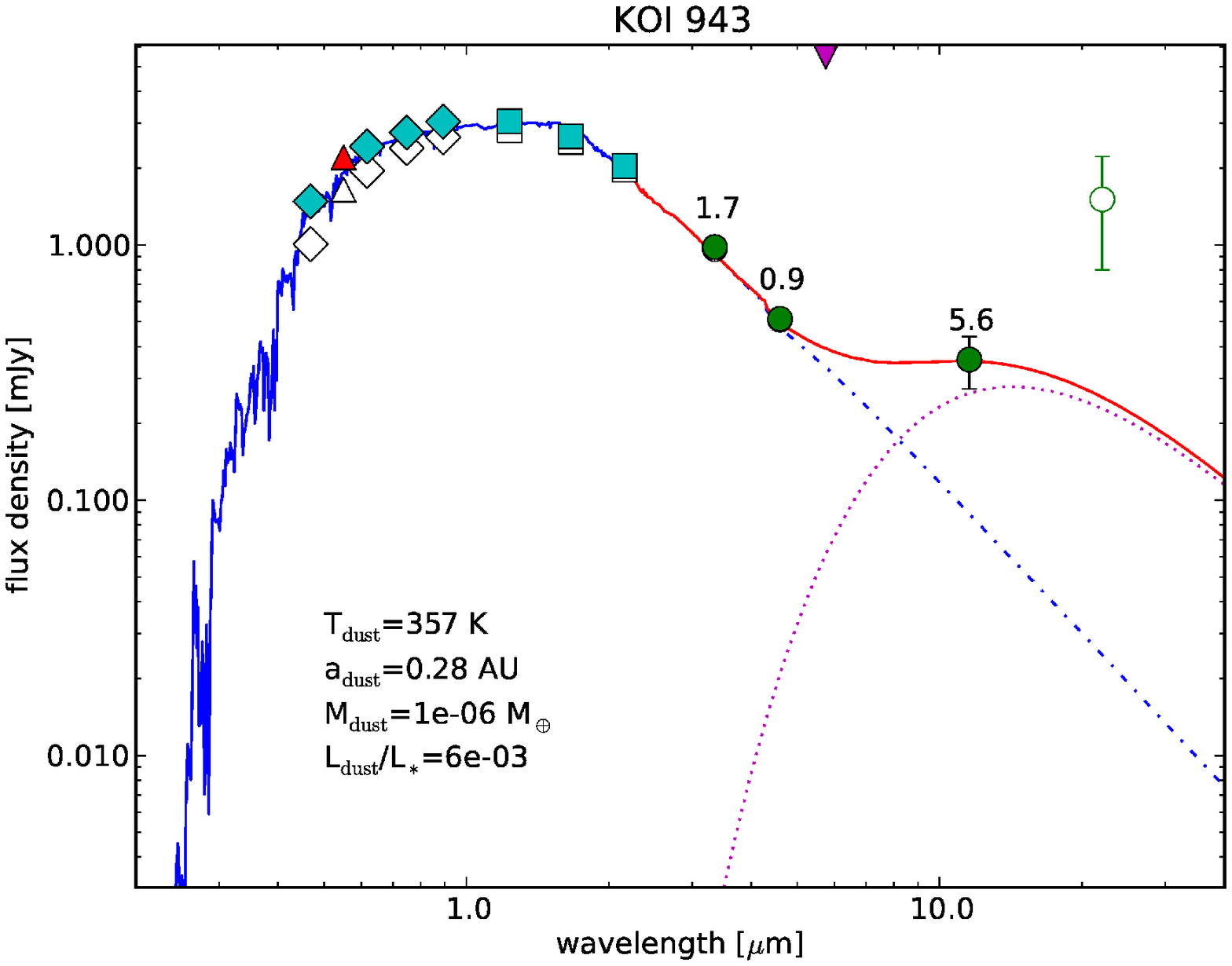}
\caption{
Stars with warm excesses, at orbital distances similar to that of the known planets.
See captions for Figures~\ref{fig:noex} and \ref{fig:sed1099}.
Open green triangles show upper limits on the W4 data, while open circles show
W4 data with SNR$<$3.
}
\label{fig:warmseds}
\end{figure}

Because of its wavelength range, WISE is most effective at detecting 
dust temperatures between 300-500~K, similar to the handful of warm debris disks
discussed in \citet{Wyattetal2007}.
Figures~\ref{fig:sed1099} and \ref{fig:warmseds} show the five systems that 
possess excesses consistent with emission by warm dust.  
Not surprisingly, due to the low photospheric flux predicted at long wavelengths,
none of these systems have W4 detections with SNR$>$3.
The low SNR datapoints and upper limits that are plotted (but not included in our fit) give tantalizing
hints of what the upcoming full WISE data release could reveal about the dust locations
in these systems.
The W4 datapoints may reveal that these systems are actually similar to the 
cool dust systems in dust mass and temperature (see Section~\ref{sec:DDs:cool}).
Only one of these five systems (KOI~1099) {\it requires} dust warmer than $>$300~K (see discussion below); 
in the other four systems we may be seeing the Wein tail of a cooler, more massive dust ring.

With the currently available WISE data, we find that
these disks are plausibly hotter than the majority of known debris disk systems, 
and our best fitting models place dust
on orbits of a few tenths of an AU.
This is perhaps not surprising, as it is similar to the orbital radii of the known planets
in these systems, so the dust may be co-located with several Earth masses
of planetary material. 
Most systems in the literature known to host both a debris disk and one or more exoplanets have a large
separation (tens of AU) between the orbit of the planet(s) and the debris disk
\citep{Brydenetal2009,DodsonRobinsonetal2011}.
One of the five systems in our sample with warm excesses has the modeled dust ring 
interior to the known planet (Figure~\ref{fig:sed1099}), with four exterior (Figure~\ref{fig:warmseds}).
All planets are within a few tenths of an AU of the modeled debris disk,
which may allow dynamical interactions between the 
planet and the disk.

The G5 star KOI~1099, shown in Figure~\ref{fig:sed1099}, is our most promising candidate to host a debris disk, 
appearing to have excesses in all WISE bands.
The weak excess in W1 combined with highly significant excesses
in both the W2 and W3 bands demands dust emission of $\sim$500~K using our nominal model.
An excellent match to all photometry between 0.5-12~$\mu$m is provided by a confined 
dust ring at 0.14~AU, although more complex distributions are certainly possible. 
The 1~$\mu$m modified-blackbody alternate model results in a cooler temperature ($\sim$400~K) and 
larger orbital radius for the dust ring (0.22~AU).
Either model results in a ring is inside the 0.57~AU orbit of the known Neptune-mass planet (candidate),
and may point to inspiral of dust past the planet from a more distant dust source.
The W4 flux has SNR$<$3 and so was not used in our fit, but if future WISE data releases 
confirm the $\sim$1.5~mJy flux, this points to an even greater dust mass, much of which 
would be at larger distances from the star.
Using the low SNR W4 datapoint along with the other WISE bands would not allow a single temperature
fit, simply indicating that this system contains dust that is producing significant emission 
at more than one orbital distance.

The other four warm systems, shown in Figure~\ref{fig:warmseds}, do not have formally 
significant W1 and W2 excesses.
It is possible that these weak W1 and W2 excesses should be ignored, 
in which case the W3 excess is the start of a blackbody of a much more massive but colder disk.
We evaluate this explicitly for the KOI~943 system, where there is a low-SNR W4 datapoint
rather than just an upper limit.

KOI~469 is a G0 type star with a super-Neptune at 0.095~AU.
Because only the W3 datapoint is off the photosphere (with only an upper limit in W4),
for our best-fitting dust ring we fixed the temperature by eye and fit the
data by scaling the mass of the ring.
This gives a dust ring at 0.36~AU, outside the orbit of the known planet.

KOI~559 is a K0 type star with a $\sim$2~$\Mearth$ planet at 0.05~AU.
There is a very weak excess in W1 and W2, with a strong excess at W3 and 
only a W4 upper limit.
The best-fitting dust ring sits at 0.23~AU, well outside this hot super-Earth's orbit.

The G5 star KOI~871 has a Saturn-mass planet at 0.1~AU.
The WISE data is similar to KOI~559, with a very weak excess in W1 and W2,
a strong excess in W3, and a W4 upper limit.
The best fitting dust ring sits at 0.27~AU, again outside the planet's orbit.

The K0 star
KOI~943 follows the usual pattern of weak excesses in W1 and W2, with a strong excess in W3.
The best-fitting dust ring sits at 0.28~AU, outside the orbit of KOI~943's super-Earth
at 0.05~AU.
The low-SNR W4 datapoint hints at a more massive, more distant dust dust ring 
that will perhaps be confirmed by the upcoming full WISE data release.
Using the nominal 1.5 mJy W4 flux in the fit pulls down the temperature to about 150~K, corresponding 
to a ring at 1.5~AU with a mass of 10$^{-4}$~$\Mearth$, similar to the 
cool dust systems described in Section~\ref{sec:DDs:cool}.

The dust masses required for all of these systems are fairly low: $\sim$10$^{-7}$-10$^{-6}$~$\Mearth$,
approximately equivalent mass to grinding a 10~km asteroid into 10~$\mu$m dust grains.

\subsection{Cool Dust Systems ($\sim$100~K)} \label{sec:DDs:cool}

\begin{figure}[h]
\centering
\includegraphics[scale=0.4]{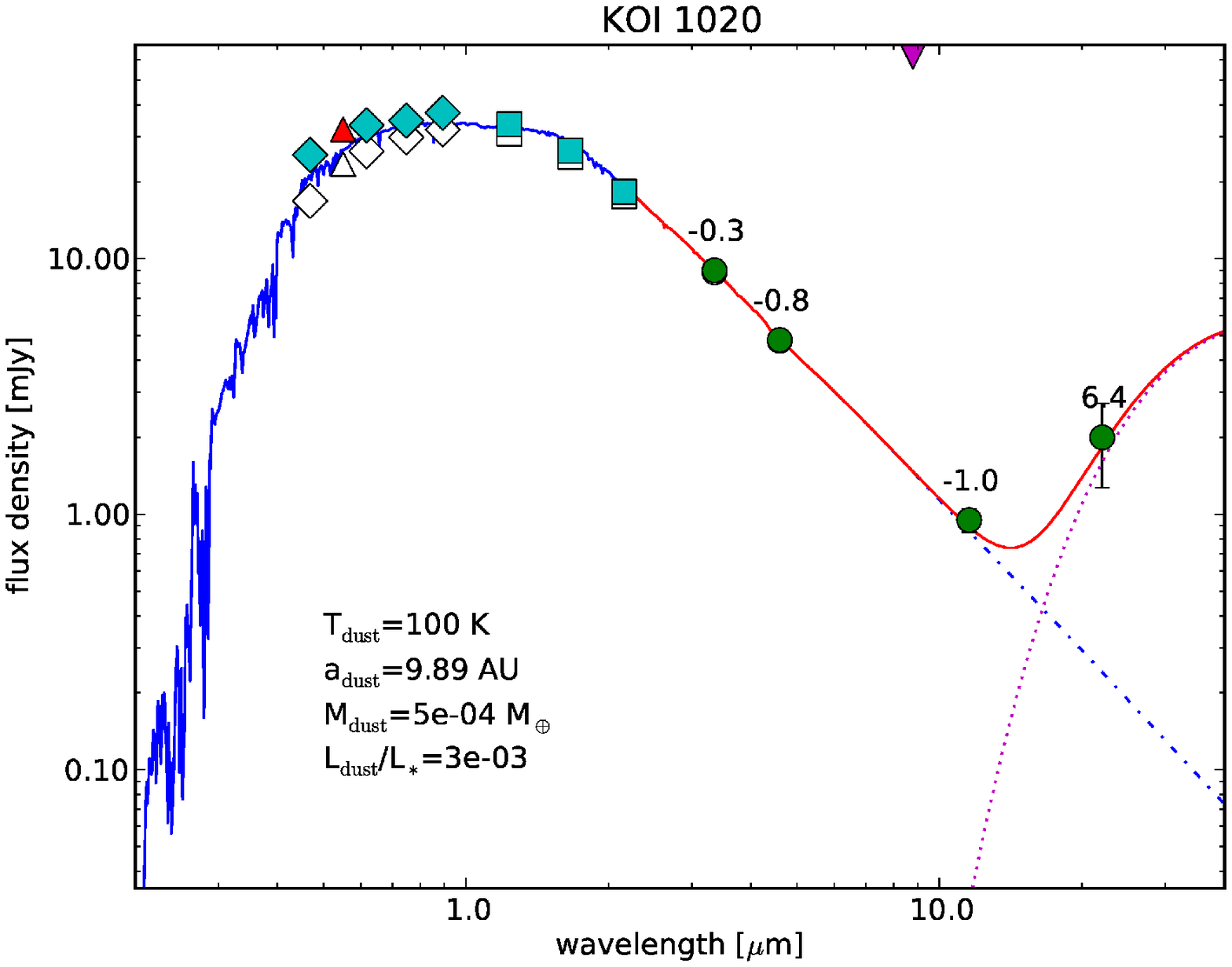} \includegraphics[scale=0.4]{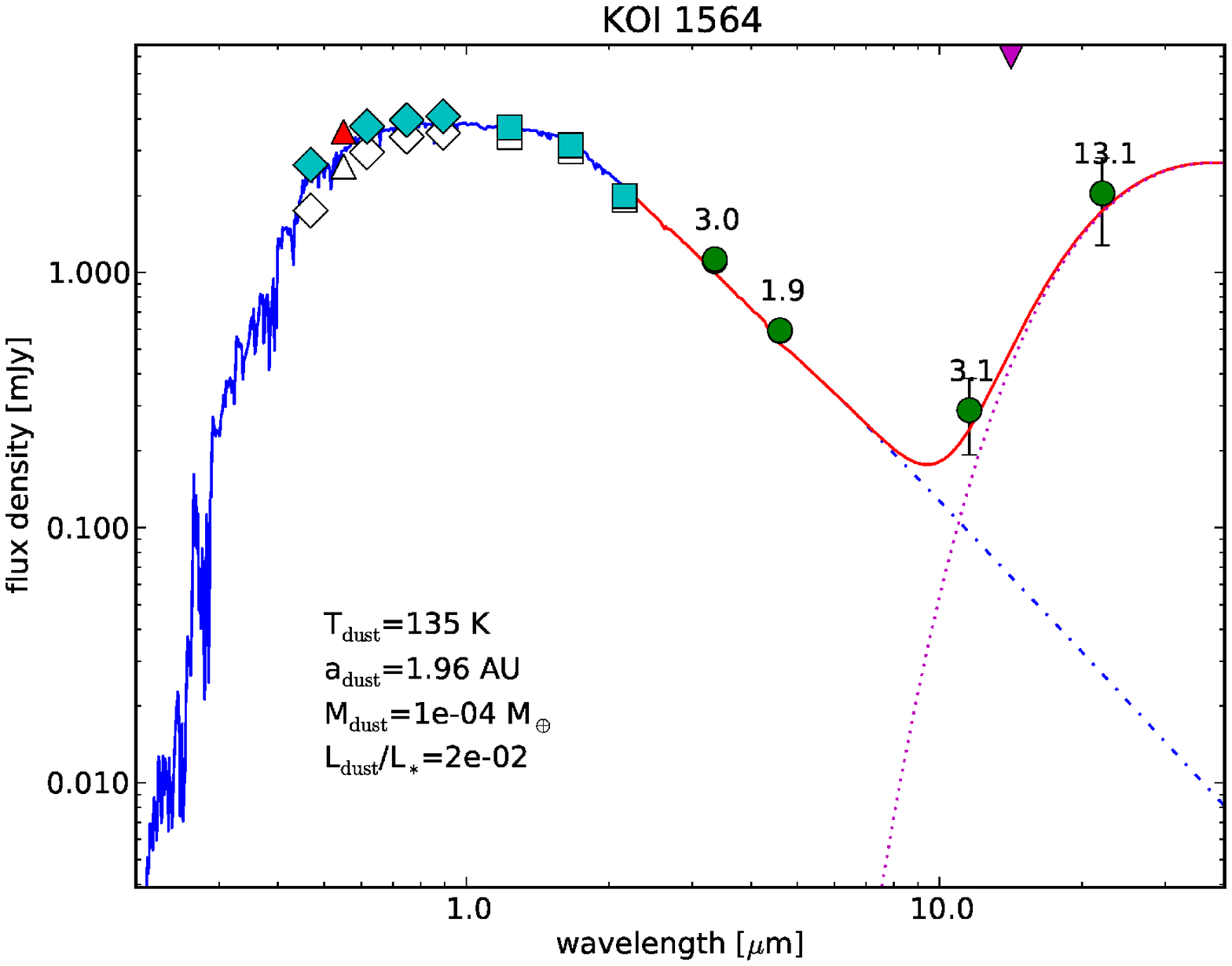}
\caption{
Stars with cool excesses ($\sim$100~K).
See captions for Figures~\ref{fig:noex} and \ref{fig:sed1099}.
}
\label{fig:coolseds}
\end{figure}

These systems (Figure~\ref{fig:coolseds}) more closely resemble classical debris disk systems that have been studied
extensively at longer wavelengths by Spitzer, with dust located at several AU,
well outside the orbits of the known planets \citep{Brydenetal2009,DodsonRobinsonetal2011}.
We find two stars in our sample with excesses that are consistent with blackbody peaks at
the W4 effective wavelength of 22~$\mu$m or longer, and strongly significant W4 detections.

KOI~1020 is a G3 star, and is the only star which has a strong W4 excess and yet no W1-W3 excess.
This would require the dust to be quite cool ($\lesssim$~100~K).  
Due to the unconstrained temperature fit, we fixed T$_{\rm dust}$ at 100~K, but it could be colder
if a more massive disk is present.
Even the 100~K temperature places the dust at about 10~AU, far outside the 
orbit of its hot Jupiter at 0.3~AU.

KOI~1564 is a G4 star, and has weak excesses in each of the 3 shortest wavelength bands, 
with a large excess in W4.  
Due to the large error bars on the W3 and W4 data, there are quite a range of temperatures
that fit, but our best dust model has an orbital distance of 2.0~AU.
KOI~1564's known Neptune-mass planet orbits well inside this, at 0.28~AU.
The future improvement in the error bars provided by the next WISE data release will certainly
better constrain this belt.

Both of these stars require high dust masses ($\sim$10$^{-4}$~$\Mearth$)
to produce the large excesses observed in W4,
similar to the masses calculated for Spitzer surveys of solar-type stars
at 24 and 70~$\mu$m \citep{Brydenetal2006,Hillenbrandetal2008},
although KOI~1564's disk may be among the hottest known with this mass or larger.

\subsection{Hot Dust ($\sim$1000~K) in a Multiplanet System} \label{sec:DDs:hot}

\begin{figure}[h]
\centering
\includegraphics[scale=0.6]{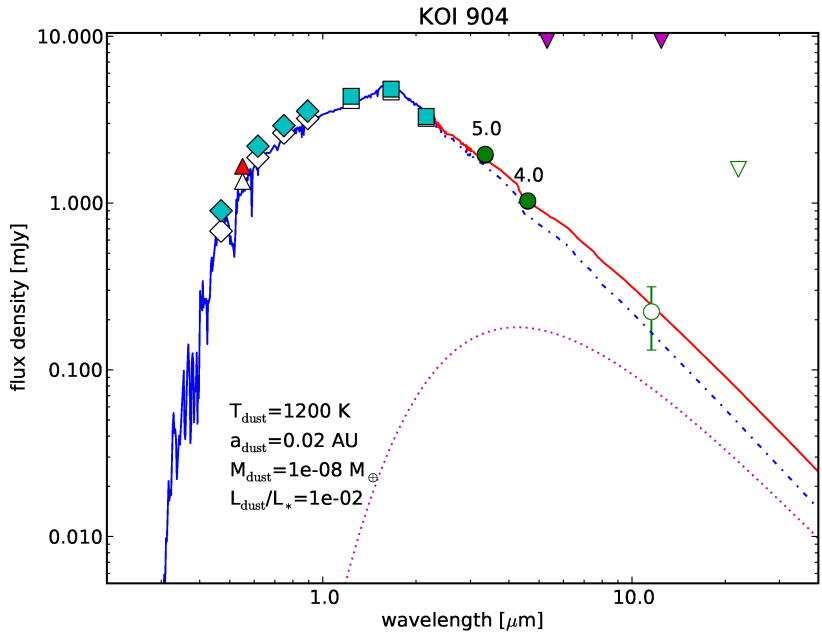}
\caption{
KOI~904: the only star with a strong excess in both W1 and W2,
and the only multiplanet system in this sample
to show an excess.
See captions for Figure~\ref{fig:noex}, \ref{fig:sed1099} and \ref{fig:warmseds}.
}
\label{fig:sed904}
\end{figure}

The K5 star KOI~904 (Figure~\ref{fig:sed904}) is the only multiplanet system to show an excess, as well as the only
system with signficant excess in the shortest wavelength band (3.4~$\mu$m).  
This alone requires extremely hot dust.
Both of the planets are super-Earths, one at 0.03~AU and one at 0.16~AU,
indicating that there is at least $\sim$15~$\Mearth$ of 
material near the star.
Having two planets is significant, because it greatly reduces the possibility
of a false positive exoplanet detection \citep{RagozzineHolman2010,LissauerNature2011}.
Currently, there is not a detection in the W3 or W4 bands, 
only a low-SNR W3 point and a W4 upper limit,
so the dust model is not well constrained.  
For lack of any other guidance on the temperature, we fixed it so the blackbody curve
peaks near the W2 effective wavelength and found an acceptable fit.  
This results in very hot dust, close to the sublimation temperature of silicates.
This exact temperature is uncertain, but must be quite hot ($>$1000~K) in order to fit 
both the W1 and W2 datapoints.
But with currently only two datapoints, more detailed dust modelling is unwarranted.

This remains our most insecure excess; there is a small mismatch between the SDSS 
and 2MASS photometry which could be the result of stellar variability 
(although the star is not flagged as variable in the WISE database).
This is among the cooler stars in the sample, and so there is the possibility 
of poor stellar atmosphere modeling, as for the M stars.  
However, other K5 and later stars appear in our sample without showing signs of 
excess in the WISE data (see Figure~\ref{fig:datahistos}).
The very hot dust is also a bit worrying, though it is not very different from the 
equilibrium temperature of the innermost known planet in this system (960~K),
and dust temperatures in the range 500-1200~K have also been detected around white dwarfs
\citep{Chuetal2011}.

If real,
this very hot dust cannot be produced by planetesimals in orbit at this distance;
as discussed in more detail in Section~\ref{sec:dust}, both the collisional timescale
and the Poynting-Robertson drag time guarantee that particles cannot stay in such close
orbits to the star for more than hundreds of years.

Only a very small dust mass ($\sim$10$^{-8}$~$\Mearth$) is required in this system 
because of the extremely hot temperatures.
Dust this hot has only been observed with optical interferometers, and may actually
be fairly common around solar-type stars \citep{Defrereetal2011}, but is difficult to observe.
Excesses this hot may have been missed by Spitzer spectroscopic surveys, which assumed
no excess was present at shorter wavelengths if no slope other than Rayleigh-Jeans was 
found in the short wavelength data \citep{Beichmanetal2006,Lawleretal2009}.
This excess is perhaps similar to that found for HD~23514 \citep{Rheeetal2008}, 
which was also detected at short wavelengths only.  
The full WISE data release should provide a high SNR W3 datapoint;
if it remains above the photosphere, the case for a tenuous, but very hot
ring of small grains near the planets will be strengthened.

We note that as this paper went to press, the February 2012 Kepler data 
release \citep{Batalhaetal2012}
provided 3 additional planets in this system, bringing the total number
of planet candidates around KOI~904 to five, all within 0.2~AU.

\section{Dust Production} \label{sec:dust}

For the systems with abundant dust interior to 1~AU,
it is difficult to produce viable steady-state models with 
such a large mass of small grains remaining
so close to their stars.
The PR drag timescale for dust to spiral into the star is extremely short at these close distances:
\begin{equation}
t_{\rm PR} = 200~{\rm yr} 
\left(\frac{r_{\rm grain}}{10~\rm \mu m}\right)
\left(\frac{a}{1~\rm AU}\right)^2
\end{equation}
for 3~g~cm$^{-3}$ grains \citep{Burnsetal1979},
giving only $\sim$10~years for 10~$\mu$m grains at 0.25~AU
(a typical orbital distance in Table~\ref{tab:DDs}).
Inspiraling grains would not ultimately reach the star in any case, as they should
vaporize at $\sim$1000-1500~K (depending on composition) once they get close enough for temperatures to 
rise this high.
But even before reaching this sublimation temperature, the collisional lifetime
may drop below the orbital timescale and dust will quickly be smashed into pieces
smaller than the blowout limit.

This timescale ($t_{\rm coll}$) for 
inter-grain collisions to produce fragments smaller than the blowout limit 
is very short at these orbital radii and number densities.
For a completely coupled narrow ring of width 2$ae$ and height 2$ai$
where all grains have the same size $r_{\rm grain}$, semimajor axis $a$,
eccentricity $e$, and inclination $i$,
$t_{\rm coll}$ can be approximated by the inverse of the collision frequency, $n \sigma v$,
where $n$ is the number density of particles, $\sigma$ is the cross-sectional area of one
particle, and $v$ is the relative speed between particles.
This gives the scaling:
\begin{equation}
t_{\rm coll} \sim
0.3~{\rm yr} \left(\frac{e}{0.1}\right)
\left(\frac{\rho}{3~\rm g~cm^{-3}}\right) 
\left(\frac{10^{-6} \Mearth}{M_{\rm grain}}\right)  
\left(\frac{r_{\rm grain}}{10~\mu \rm m}\right)
\left(\frac{a}{0.25~\rm AU}\right)^{3.5}
\end{equation}
where $\rho$ is the density of a dust grain, $M_{\rm grain}$ is the total mass 
of the dust grains, and the approximation has been made that $e \approx i$ (where $i$ is in radians).
With $t_{\rm coll}=0.3$~yr for our typical warm systems,
a steady state requires replenishing $\sim$10$^{-6}$~$\Mearth$ (or more) of dust
every year, consuming $>$1000~$\Mearth$ of material in 1~Gyr,
which is unreasonably large.

Unrealistic mass-consumption rates implied by such steady-state models are worrying, 
and have been noted for several main-sequence, warm debris disk systems.
The fractional luminosities L$_{\rm dust}$/L$_*$ we measure for Kepler systems 
with excesses also greatly exceed the 
maximum predicted by the steady-state collisional model of \citet{Wyattetal2007}
given these stars' mature ages.
The common logical conclusion is that dust masses at this level 
have to be transient.
Several theories have been proposed to explain such systems, including 
comet swarms, a ``supercomet'', or a recent collision \citep[HD~69830;][]{Beichman69830}, 
or a late-heavy bombardment-style dynamical instability forcing many icy
bodies onto highly eccentric orbits, bringing them in 
close to the star where they collide or sublimate 
\citep[$\eta$~Corvi;][]{Wyattetal2007,Lisseetal2011}.

Another possible explanation for these massive debris disks may be that these stars 
are all young ($<$100~Myr) and still in the process of clearing their protoplanetary disks.
This hypothesis was offered for three of the seven stars presented by 
\citet{Wyattetal2007} with
disks far too massive to be maintained by a steady-state collisional cascade.
We inspected Palomar Sky Survey images and found no evidence that any of 
our sample stars are associated with the open clusters in the Kepler field,
reducing the probability that this explains these systems.

A further option is that the dust may be impact ejecta from 
asteroids or comets recently striking the extant exoplanets, similar to the production of
giant planet rings in our Solar System \citep{Burnsetal1999};
however this is challenging given the super-Earths' large escape speeds, 
even if impact speeds are also high.

What is clear is that these dust grains cannot be co-located 
with a planetesimal belt that is producing them.
A massive asteroid belt at 0.2~AU or closer is an
unsatisfactory dust source because $t_{\rm coll}$ for the planetesimals is $\ll1$~Gyr
\citep{Wyattetal2007,Moldovanetal2010} and the source would now be gone.
Dust also cannot simply spiral in to these close distances from a more distant source
in steady-state.
The high optical depths that we calculate must be present in these systems
(ranging from $\tau\sim0.01-0.5$)
are several orders of magnitude higher than the maximum allowed \citep{Wyatt2005}
in a steady-state inspiral scenario,
meaning that the dust present in these quantities would destroy itself collisionally before
PR drag has a chance to move it any significant distance.

Given the large dust masses and the small fraction of stars that possess
these excesses, 
episodic dust production mechanisms provide the most attractive explanation.
Dust could be produced by a catastrophic collision between asteroids located further from
the star, and we are seeing the dust as it spirals in toward the star.
However, the probability of occurance required by this theory may not be high enough to 
explain the observed fraction of such systems.
Dynamical instabilities forcing many bodies from the outer reaches of these solar systems
onto eccentric orbits so they collide close to the star, perhaps similar in style to 
the Late-Heavy Bombardment in our Solar System, might be seen as an attractive theory.
However, the fraction of systems observed to have 24~$\mu$m excesses observed by Spitzer
appears to be much too high given the short lifetime of such an event \citep{Boothetal2009}.

\section{Discussion}  \label{sec:conc}

We find IR excesses around eight stars that also host planet candidates.
The masses and luminosities of these excesses are all too large to be explained by 
a steady-state collisional cascade.
Given that the WISE field coverage was not uniform in depth, and so many of 
these relatively faint stars suffered from contamination by nearby bright stars, 
it is challenging to say anything robust about the fraction 
of these planet-hosting stars that have excesses in our sample.
Out of the 186 stars with W3 data of SNR$>$3, five have significant $>$5~$\delta$
excesses, or $\sim$3\%.  
At first glance, this fraction appears similar to the 2\% of
solar-type stars with warm excesses \citep{Wyattetal2007,Lawleretal2009}.

However, the fractional luminosities of these Kepler systems with excesses are
much higher than most known debris disks \citep[Figure~\ref{fig:dustlocations} and][]{Mooretal2006}.
Because of the larger distances and thus lower apparent brightness of these Kepler systems
as compared to most known debris disk systems, only the brightest debris disks would be 
detected by WISE.  
Presumably the eight systems presented here are the very highest luminosity debris disks in the sample,
and many more fainter disks are present around these planet-hosting stars.
This would indicate that the fraction of stars possessing debris disks and Kepler-detected
exoplanets is actually much higher than the fraction of field stars with debris disks.

Surveys with Spitzer \citep{Brydenetal2009,DodsonRobinsonetal2011} 
have found that hosting a radial velocity-detected exoplanet does not make a star more likely to host a debris disk,
and a recent WISE survey of exoplanets discovered by ground-based transit searches 
\citep{Krivovetal2011} has found that 
the fraction of these stars ($\sim$2\%) to host warm debris disks (excesses in W3 and W4)
is similar to that found by \citet{Wyattetal2007}.

The Kepler systems studied here are different from other 
disk-bearing exoplanet systems, 
whose planets are primarily discovered by the radial-velocity (RV) technique,
and by ground-based transit searches.
Exoplanets from the latter two techniques are mostly large and very close to their stars, 
with an average mass of $\sim$2 Jupiter masses
and $<$20\% known to be in multiplanet systems \citep{Wrightetal2011}.
The recently announced 1200 Kepler exoplanet candidates \citep{Boruckietal2011} 
are in a different regime,
with about 1/3 in multiplanet systems, 
and having many super-Earths (typically 2-6 Earth radii) well inside 1~AU.
Among our excess systems, all but two of the planets are Neptune mass or smaller;
only one is massive enough to be a hot Jupiter.
The resulting correlation we find between a star hosting both small planets and 
a debris disk inside a few AU agrees with the planetary formation and dynamical evolution
models presented by \citet{Raymondetal2012}.
This makes sense in a picture where migrating Jupiters clean out the 
terrestrial systems they pass through.

Explaining the large dust masses required to produce these excesses remains a huge challenge.
Steady-state theories appear to fail completely, as the mass consumption rates
required would involve destroying thousands of Earth masses of planetary material
over the lifetime of these stars.
Episodic events such as catastrophic collisions between bodies may provide an explanation,
but these events must last long enough and occur frequently enough to result in $\sim$3\% of
solar-type stars in this state at any given moment.
Much theoretical work needs to be done here.

The upcoming WISE full-sky data release will access additional systems,
decrease the error bars on existing measurements, and provide
longer-wavelength measurements for some of these systems to see if there is dust at larger orbital distances.
Spectral data on these candidates would be invaluable, but the host stars are faint.
Future Kepler data releases will contain planets on 
more distant orbits, perhaps yielding planets in closer proximity to these new debris disks.

\acknowledgments {
The authors wish to thank Angelle Tanner, Ryan Goldsbury, Harvey Richer, Joe Burns,
Casey Lisse, Alexander Krivov, Kate Su, and Geoff Bryden for helpful discussions, 
and an anonymous referee whose comments greatly improved our interpretation of the data.
The authors wish to thank both the Kepler and WISE science teams not only for carrying
out incredibly successful missions, but for making their data publicly available.

This paper includes data collected by the Kepler mission. 
Funding for the Kepler mission is provided by the NASA Science Mission directorate.
Some of the data presented in this paper were obtained from the Multimission Archive 
at the Space Telescope Science Institute (MAST). 
STScI is operated by the Association of Universities for Research in Astronomy, Inc., 
under NASA contract NAS5-26555. 
Support for MAST for non-HST data is provided by the NASA Office of Space Science via grant 
NNX09AF08G and by other grants and contracts.
This publication makes use of data products from the Wide-field Infrared Survey Explorer, 
which is a joint project of the University of California, Los Angeles, and the 
Jet Propulsion Laboratory/California Institute of Technology, funded by the National 
Aeronautics and Space Administration. 
This publication also makes use of data products from the Two Micron All Sky Survey, 
which is a joint project of the University of Massachusetts and the Infrared Processing 
and Analysis Center/California Institute of Technology, funded by the National Aeronautics 
and Space Administration and the National Science Foundation.
This research has made use of the NASA/ IPAC Infrared Science Archive, which is operated by 
the Jet Propulsion Laboratory, California Institute of Technology, under contract with the 
National Aeronautics and Space Administration.

This paper is dedicated to Sylvia Maria Bongarzone Lawler who was born December 23, 2011.
}


\emph{Note added in proof:} During the publication process the WISE full data release became available, which updates fluxes and upper limits for some of the systems analyzed here, and presents measurements for additional Kepler targets.  In particular, the W3 and W4 measurements in the KOI 1099 system have become upper limits.  Reanalysis of the revised and expanded dataset is presented in a forthcoming paper.

\end{document}